\documentclass{article}
\usepackage[onecolumn]{emulateapj,epsfig}
\long\def\***#1{{\scshape ***#1***}}

\makeatletter

\newenvironment{inlinefigure}{%
\def\@captype{figure}%
\noindent\begin{minipage}{0.999\linewidth}\begin{center}}
{\end{center}\end{minipage}\smallskip}
\makeatother

\lefthead{SONGAILA \& COWIE}
\righthead{APPROACHING REIONIZATION}

\begin{document}

\submitted{Astronomical Journal, in press (May 2002)}
\title{Approaching Reionization: The Evolution of the Lyman Alpha Forest from
Redshifts Four to Six}
\author{Antoinette Songaila\altaffilmark{1} \& Lennox L. Cowie\altaffilmark{1}}
\affil{Institute for Astronomy, University of Hawaii, 2680 Woodlawn Drive,
  Honolulu, HI 96822\\}

\altaffiltext{1}{Visiting astronomer, W. M. Keck
  Observatory, jointly operated by the California Institute of Technology and
  the University of California.}

\slugcomment{Astronomical Journal, in press (May 2002)}

\begin{abstract}
We analyze the Ly$\alpha$\ forest properties of a sample of 15 high redshift
quasars lying between $z = 4.42$\ and $z = 5.75$, using high signal-to-noise
spectra obtained with ESI on the Keck~II 10~m telescope.  The distribution of
transmissions in the Ly$\alpha$\ region in this redshift range is shown to be
consistent with that found in lambda cold dark matter simulations with a
smoothly evolving ionization rate as a function of redshift.  The
extrapolation of the ionizing flux to $z = 6.05$\ lies a factor of two higher
than a $2~\sigma$\ upper limit placed by Cen \& McDonald (2001) at this
redshift, based on the Becker et al.\ (2001) spectra of the $z = 6.28$\ quasar
SDSS~1030+0524.  However, the data are also consistent with models in which
there is substantial variation of the ionization parameter about the mean
value, and in this case, dark gaps such as those seen by Becker et al.\ become
much more likely.  We conclude that further data are needed before we can be
sure that we have seen the epoch of reionization.  We also summarize the
damped Ly$\alpha$\ systems seen in these quasar lines of sight and measure
their metallicities and kinematic properties.  We argue that the mean DLA
metallicity has dropped substantially by $z = 5$\ compared with its value at
$z < 4$.

\end{abstract}


\section{Introduction} \label{intro}
The epoch of reionization marks a fundamental change in the intergalactic gas
in which, as the ionized underdense regions merge and overdense regions become
exposed to the combined ionizing flux of all the galaxies and AGN, the gas
transits from being predominantly neutral to being predominantly ionized over
a relatively short redshift interval (e.g.\ Razoumov et al.\ 2001, who find
$\Delta z \sim 0.3$).  As surveys for high $z$\ objects have proceeded rapidly
in the last two years, we are approaching redshifts at which the effects of
reionization on the IGM might be expected to be becoming apparent.  The
current highest redshift object is the $z = 6.56$\ Ly$\alpha$\ emitter found
by Hu et al.\ (2001) who argue that this object suggests that
reionization lies beyond $z = 6.6$.

Of more interest for studying the {\it details\/} of the intergalactic gas are
the extremely luminous quasars being found by the Sloan digital Sky Survey
(SDSS) out to a redshift of 6.28 at the present time (e.g. Fan et al.\ 2001a
and references therein).  In recent papers, Djorgovski et al.\ (2001) and
Becker et al.\ (2001) (see also Pentericci et al.\ 2001 and Fan et al.\ 2001b)
have presented evidence from medium-resolution spectra of the highest redshift
SDSS quasars that they claim supports the low redshift edge of reionization at
$z \sim 6$.  There are a number of
limitations in both analyses, largely stemming from the small redshift
baselines in both cases.  Djorgovski et al.\ used high-quality KeckI LRIS and
KeckII ESI spectra of the $z_{\rm em} = 5.75$\ quasar SDSS~1044$-$0125 to 
suggest that they had seen the onset of the 'trailing edge
of the reionization era' at $z \sim 5.5$, based on the presence of opacity
gaps in their spectrum blueward of Ly$\alpha$\ emission.  
Becker et al.\ presented relatively low signal-to-noise spectra of the four
highest redshift SDSS quasars and concluded that there is rapid evolution of
the mean absorption at these redshifts, including SDSS~1030+0524 with $z_{\rm
em} = 6.28$, in which they claimed to see a complete Gunn-Peterson trough at
the longest wavelengths observed.

As Becker et al.\ correctly point out, the appearance of a single
Gunn-Peterson trough over a short wavelength interval is not in itself
conclusive evidence that the epoch of reionization has been reached since only
a small fraction of neutral hydrogen can cause this.  The crucial diagnostic
of reionization is the evolution of the ionizing flux (cf.\ Fan et al.\ 2001b;
McDonald \& Miralda-Escud\'e 2001), and for these purposes, a longer redshift
baseline is important to deconvolve the effects of the structural thickening
of the forest from the ionizing radiation evolution.  In this paper we present
such a discussion based on ESI spectra of 15 quasars with $4.42 < z_{\rm em} <
5.745$, all of which have a quality comparable to or better than Djorgovski et
al.`s spectrum of SDSS~1044$-$0125, and all of which have highly optimised sky
subtraction, a crucial point at the low transmitted flux levels characteristic
of these redshifts.  The transmitted flux presents a smooth evolution to a
redshift of 6 and the comoving emission rate per unit volume of ionizing
photons is constant or only very slightly declining at these redshifts.  If
extrapolated, is still sufficient to maintain a highly ionized IGM to $z =
6.6$.  

This leaves the longest redshift ($z = 6.05$) point in SDSS~1030+0524 as
the single strong piece of evidence for the epoch of reionization lying at $z
= 6.1$.  Cen \& McDonald (2001) have used a $\Lambda$-dominated cold dark
matter (LCDM) model (which provides a good fit to our present data) to analyze
the Ly$\alpha$\ and Ly$\beta$\ absorption of this point, concluding that the
$2~\sigma$\ upper limit on the ionization rate at this redshift is about half
of that which we would deduce from an extrapolation of our lower redshift
data.  An abrupt drop would be consistent with the Becker et al.\ and Fan et
al.\ interpretations.  However, as we show, the lower redshift data is also
consistent with models in which there is substantial dispersion in the
ionization parameter, or equivalently the equation of state of the gas, and
these models have a significant probablility of including dark points like
that of Becker et al.\ even for a smooth extrpolation of the mean ionization
rate.  A conclusive determination of the reionization epoch therefore requires
a much larger sample of data at the highest redshifts.

\section{Background}

As was first pointed out by Shklovskij (1964), Scheuer (1965) and Gunn \&
Peterson (1965), a fully neutral intergalactic medium produces an enormous
Lyman $\alpha$\ scattering
\begin{equation}
\tau_{\rm H~I} = 3.4 \times 10^5 \, h^{-1} \, 
                 \left ( {{\Omega_m}    \over {0.35}}\right )^{-0.5} \, 
                 \left ( {{\Omega_bh^2} \over {0.0325}} \right ) \, 
                 \left ( {{1+z} \over {7}} \right )^{1.5}
\end{equation}
\medskip\noindent where, at high redshifts in the currently favored cosmology,
the local Hubble constant $H \approx \Omega_m^{1/2}\, (1+z)^{3/2}\, H_0$\
(e.g. Barkana 2001).  Here $h = H_0/(100~{\rm km\ s^{-1}\ Mpc^{-1}})$,
$\Omega_m$\ is the local mass density and $\Omega_b$\ is the baryon density.
Even when the effects of structure are allowed for at these redshifts, the
minimum density 'void' regions with baryon densities $\Delta \equiv
\rho_b/\langle\rho_b\rangle \sim 0.1$\ would have optical depths in excess of
$10^4$.  A predominantly neutral IGM would therefore produce a totally black
Ly$\alpha$\ forest region.  This is quite unlike the situation seen in the
spectra of quasars to $z = 6.2$\ where, as we discuss in detail below, there
are still substantial portions of the spectrum in which there is transmitted
flux.

In actuality the lower density regions of the IGM will become ionized prior to
the point at which the ionized regions merge and the bulk of the mass of the
the IGM begins to ionize, which we would consider to be the redshift of
reionization (see Miralda-Escud\'e, Haehnelt \& Rees 2000 for a detailed
description).  The transmission through the underdense ionized regions results
in some observable flux.  However, even in this limit the radiation damping
wings of the neutral hydrogen regions can completely black out the region.  A
spectrum with portions that are still transmitting in the Lyman $\alpha$\
forest region can correspond to a period near reionization only if the dark
gaps produced by the substantial amounts of neutral hydrogen in the high
overdensity regions are sufficiently separated.  This in turn means that the
dark gaps must contain high column densities of neutral hydrogen and be very
wide (cf. Barkana 2001).

A simple toy model serves to illustrate this point.  Suppose the baryons are
substantially neutral (fraction $f_N$) by mass and that all the neutral
material is concentrated in a 'picket fence' with the separation between the
regions corresponding to a redshift interval $\Delta z$\ or rest-frame
wavelength separation $\Delta \lambda = \lambda_{\alpha}\, \Delta z / (1+z)$.
This model neglects the neutral hydrogen opacity contributions from the
ionized regions, and so minimizes the 'darkness' of the spectrum.  The column
density of each spike in the picket fence is now
\begin{equation}
N({\rm H~I}) = \left ( {{c\rho_c} \over {H_0 m}}\right ) \, \Omega_m^{-0.5} \,
(1+z)^{1.5} \, f_N\,\Omega_b \, {{\Delta \lambda} \over {\lambda_{\alpha}}}
\end{equation}
\smallskip\noindent
where $\rho_c$\ is the closure density, $m$ the mass per hydrogen atom and
$\lambda_{\alpha}$\ is the Lyman $\alpha$\ wavelength.  Neutral hydrogen
regions have a damped Lyman $\alpha$\ equivalent width of
\begin{equation}
W_{\alpha} = 23.2 \ \left ( {{N({\rm H~I})} \over {10^{21}~{\rm
cm}^{-2}}}\right ) ^{0.5} {\rm\AA} \  \equiv \   23.2\,N_{21}^{0.5} \ {\rm\AA}
\end{equation}
\smallskip\noindent
and produce dark regions 
\begin{equation}
\Delta \lambda_{2.5} = 8.3\,N_{21}^{0.5}\ {\rm\AA} \quad {\rm and} \quad \Delta
\lambda_{1.25} = 11.7\,N_{21}^{0.5}\ {\rm\AA} 
\end{equation}
\smallskip\noindent
where $\Delta \lambda_{2.5}$\ is the width of the region with $\tau > 2.5$\
and $\Delta \lambda_{1.25}$\ is that at $\tau = 1.25$.  In order to produce
regions with $\tau < 2.5$\ in the spectra, $\Delta \lambda > \Delta
\lambda_{1.25}$, which requires
\begin{equation}
N_{21}^{0.5} > 1.2 f_N\, \left( {{\Omega_m} \over {0.35}} \right )^{-0.5}\,
                         \left( {{\Omega_b h^2} \over {0.0325}} \right )\,
                         \left( {{h} \over {0.65}} \right )^{-1}\,
                         \left( {{1+z} \over {7}} \right )^{1.5}\,
\end{equation}
\smallskip\noindent
or minimum gap lengths at $\tau = 2.5$\ of
\begin{equation}
\Delta \lambda_{2.5} > 10\ {\rm\AA} \ f_N\, 
                         \left( {{\Omega_m} \over {0.35}} \right )^{-0.5}\,
                         \left( {{\Omega_b h^2} \over {0.0325}} \right )\,
                         \left( {{h} \over {0.65}} \right )^{-1}\,
                         \left( {{1+z} \over {7}} \right )^{1.5}\, .
\end{equation}
\smallskip\noindent This is an extreme lower limit: complexifying the neutral
hydrogen structure or including the opacity effect of the ionized gas will
only raise $\Delta \lambda_{2.5}$.  In essence, we must concentrate the
neutral hydrogen to produce transmission gaps, and these concentrations will
in turn produce very wide dark regions.  Thus, a key signature of the approach to
reionization is the onset of either a {\it completely black\/} spectrum or of
many {\it wide\/} dark regions.

However, distinguishing this from a highly ionized post-reionization
intergalactic medium may be extremely difficult.  At these redshifts, a
uniform ionized IGM would have a Ly$\alpha$\ optical depth
\begin{equation}
\tau_u = 14 \, \Gamma_{-12}^{-1}  \, T_4^{-0.75} \,
             \left ( {{\Omega_m} \over {0.35}} \right )^{-0.5}
             \left ( {{\Omega_bh^2} \over {0.0325}} \right )^2
             \left ( {{H_0} \over {65~{\rm km\ s^{-1}\ Mpc^{-1}}}} \right )^{-1}
             \left ( {{1+z} \over {7}} \right ) ^{4.5} 
\end{equation}
\medskip\noindent where $\Gamma_{-12}$\ is the local ionization rate produced
by the metagalactic ionizing flux in units of $10^{-12}~{\rm s}^{-1}$\ and
$T_4$\ is the gas temperature in units of $10^4$~K.  
The corresponding optical depth at Ly$\beta$\ is a combination of the direct
Ly$\beta$\ absorption and Ly$\alpha$\ absorption at the redshift
\begin{equation}
1 + z_{\beta} \equiv 
              \left ( {{\lambda_{\beta}} \over {\lambda_{\alpha}}}\right ) \,
              (1+z)
\end{equation}\smallskip\noindent
where $\lambda_{\alpha}$\ and $\lambda_{\beta}$\ are the Ly$\alpha$\ and
Ly$\beta$\ wavelengths, and is given by
\begin{equation}
\tau_{\beta_u} = 0.16 \, \tau_u(z) + \tau_u(z_{\beta}) \quad .
\end{equation}\smallskip\noindent
The two terms in equation~(9) are generally comparable, the exact ratio being
dependent on the steepness of the evolution of $\tau_u$\ with $(1+z)$.  Thus,
even in this limit, transmission again will be seen only in the most
underdense 'void' regions of the IGM, with most of the spectrum again being
black even at Ly$\beta$.  (Because of the very steep redshift dependence, this
is unlike the situation at lower redshifts, $z \sim 3$, where even the mean
density regions have significant Ly$\alpha$\ transmission.)  

The distribution
of the transmission and the mean transmitted flux depend in this high redshift
limit on the fraction of the volume occupied by regions of low density (which
in turn primarily depends on the details of the cosmological model) and on the
normalized ionization rate
\begin{equation}
g \equiv \Gamma_{-12} \, T_4^{0.75} \, 
               \left ( {{\Omega_m} \over {0.35}} \right )^{0.5} \,
               \left ( {{\Omega_b h^2} \over {0.0325}} \right )^{-2} \,
                \left ( {{H_0} \over {65~{\rm km\ s^{-1}\ Mpc^{-1}}}} \right ) 
\end{equation}\smallskip\noindent
which specifies the neutral fraction at $\Delta = 1$\ (see Mcdonald \&
Miralda-Escud\'e 2001).  For a specified cosmological model, the mean
transmitted flux in the spectrum can be used to obtain $g$\ and so the
evolution of the ionizing flux with redshift (McDonald \& Miralda-Escud\'e
2001; Cen \& McDonald 2001).  If we ignore the effects of thermal broadening
and peculiar motions, and also assume that the gas has an equation of state $T
\sim \Delta^{\gamma - 1}$, we can generalise equation (7) to obtain the
optical depth of the fluctuating medium
\begin{equation}
\tau_{\Delta} = 14\, g^{-1}\, 
                \left ( {{1+z} \over {7}} \right )^{4.5}\, \Delta^{\beta}
\end{equation}\smallskip\noindent
where $\beta = 2 - 0.75(\gamma -1)$\ and $\tau_{\Delta}$\ is the optical depth
at a point in the spectrum corresponding to a spatial position with normalized
density $\Delta$\ (e.g. Hui \& Gnedin 1997).  It can be seen immediately from
equation~(10) that transmission regions with $\tau < 1$\ will correspond to
normalized densities of less than or around $\Delta = 0.27 g^{-0.5}$\ at $z =
6$, where for simplicity we have assumed an isothermal equation of state.
Because the volume occupied by low density regions drops rapidly with
decreasing $\Delta$, the mean transmission comes from a sharply peaked region
of $\Delta$.  If we adopt (as did Fan et al.\ 2001b) the form of the volume
density distribution function given by Miralda-Escud\'e, Haehnelt \& Rees
(2000)
\begin{equation}
P(\Delta)\, d\Delta = A \, \Delta^{-b} \, 
            \exp \left ( {{-(\Delta^{-2/3} - C)^2} \over {8\delta_0^2 /9}}
                 \right ) \, d\Delta
\end{equation}\smallskip\noindent
where $b \approx 2.5$\ at
high redshift and $A$\ and $C$\ are specified by the normalization of $P$\ and
the condition $\bar{\Delta} = 1$, then ignoring the normalizing constant $C$\
term in the low density regime, the mean transmitted flux has the form
\begin{equation}
F = \int A\, \Delta^{-b}\, \exp - \left ( \tau_u\Delta^{\beta} + 
                        {{\Delta^{-4/3}} \over {8\delta_0^2 /9}} \right )
                        \, d\Delta \quad .
\end{equation}\smallskip\noindent
The exponent peaks sharply at
\begin{equation}
\Delta_0 = \left ( {{3} \over {2\tau_u \beta \delta_0^2}} 
           \right ) ^{{1} \over {\beta + 4/3}}
\end{equation}\smallskip\noindent
which, adopting $\delta_0 = 7.61 (1+z)^{-1}$\ appropriate to the LCDM model of
Miralda-Escud\'e et al.\ (1996), corresponds to 
\begin{equation}
\Delta_0 = 0.39 \, g^{0.3} \, 
           \left ( {{1+z} \over {7}} \right )^{-0.75} 
\end{equation}\smallskip\noindent
and where we have again used the isothermal equation of state for simplicity.  
We can now evaluate the integral using the method of steepest descents to give
\begin{equation}
F(z,g) =   A \, \left ( {{4\pi} \over {3\beta + 4}} \right )^{0.5}  
         \delta_0 \, \Delta_0^{5/3 - b} \,
         \exp \left ( -\left ({{3} \over {2\beta}} + {{9} \over {8}}\right )\,
              \Delta_0^{-4/3} \, \delta_0^{-2} \right )
\end{equation}\smallskip\noindent
where, for the isothermal $\beta = 2$, the exponent becomes $-5.56
g^{-0.4}((1+z)/7)^3$, so that the effect of the structure is to make the
evolution of the effective optical depth $-{\rm ln} T$\ with redshift
shallower than in the uniform case (e.g.\ Bi \& Davidsen 1996).
The dependence on the equation of state is extremely weak.
Inclusion of the $C$\ term changes the integral slightly, but 
adopting this functional form, we use the analytic equation
\begin{equation}
F(z,g) = 4.5 \, g^{-0.28} \, 
         \left ( {{1+z} \over {7}} \right )^{2.2} \,
         \exp \left ( -4.4 \, g^{-0.4} \, 
                      \left ( {{1+z} \over {7}} \right )^3 \right )
\end{equation}\smallskip\noindent
which provides an accurate representation (a maximum of 8\% deviation from $g
= 0.2$\ to $g = 3$) of the LCDM numerical models of
McDonald \& Miralda-Escud\'e (2001), Cen \& McDonald (2001) and Cen \&
McDonald (2002, in preparation) and to the exact integral obtained using
equations~(11) and (12) over the $z = 4 - 6$\ redshift range of interest.


%
%
\begin{inlinefigure}
\psfig{figure=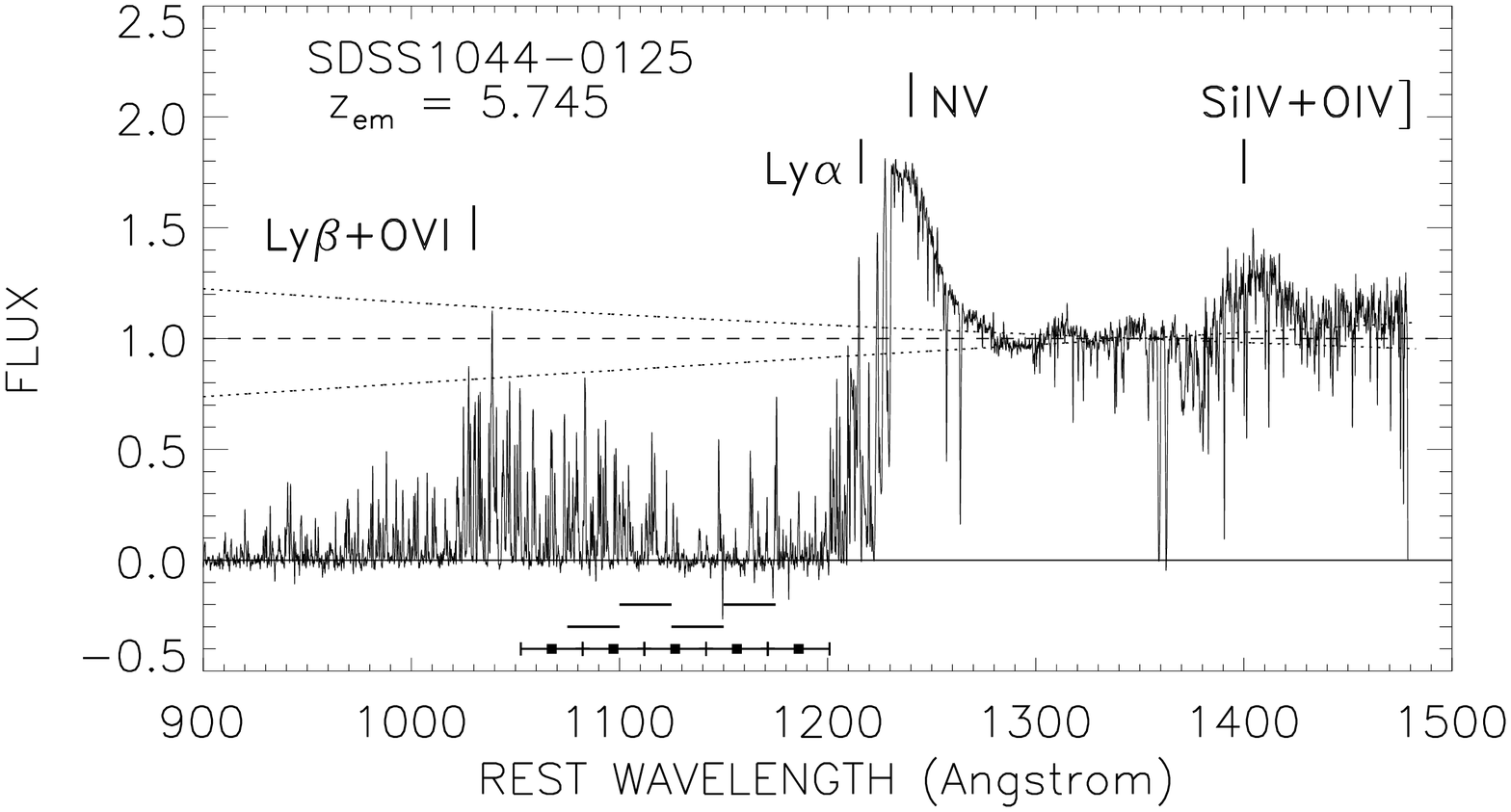,angle=90,width=7.0in}
\vspace{6pt}
\figurenum{1}
\caption{Spectrum of the quasar SDSS1044$-$0125 taken with ESI on KeckII (see
Table 1).  The emission redshift of 5.745 was adopted from Goodrich et al.\
(2001) and the positions of emission lines of Ly$\beta$+O~VI, Ly$\alpha$, N~V and
Si~IV+O~IV] are shown as short vertical lines.
The spectrum is shown smoothed to $0.25~{\rm\AA}$\ and normalized in the
region of $1330 - 1370~{\rm\AA}$\ in the rest frame to a power-law continuum
($f_\lambda$) with
slope $-1.25$ (solid line).  The dotted lines show continua with slopes $-0.75$\
and $-2$.  The short horizontal lines below the spectrum show the
positions of the wavelength bins used to calculate the transmitted fluxes in
Figures~5 and 6; they are $1075-1100~{\rm\AA}$, $1100-1125~{\rm\AA}$,
$1125-1150~{\rm\AA}$\ and $1150-1175~{\rm\AA}$\ in the rest frame.  The bins
used by Becker et al.\ (2001) are shown below that by the horizontal lines with the
centers shown by small filled squares.  Associated absorption can be seen in
the Si~IV line at wavelengths shortward of the Si~IV emission.  Based on the
corresponding C~IV, Goodrich et al.\ (2001) described this as a BAL QSO.  }
\label{fig1}
\addtolength{\baselineskip}{10pt}
\end{inlinefigure}

\section{Observations} \label{obs}

A total of 15 high redshift quasars, with $4.42 < z_{\rm em} < 5.745$, were
observed for this program (Table~1) using the Echellette Spectrograph and
Imager (ESI; Sheinis et al.\ 2000) intrument on the KeckII telescope in
echellette mode.  The resolution in this configuration is comparatively low,
$\sim 5300$\ for the $0.75^{\prime\prime}$\ slit width used, but the red
sensitivity of the instrument is high and the wavelength coverage is complete
from $4000~{\rm \AA}$\ to $10,000~{\rm \AA}$.  At the red wavelengths,
extremely high precision sky subtraction is required since the sky lines can
be more than two orders of magnitude brighter than the quasars. In order to
deal with this issue, individual half-hour exposures were made, stepped along
the slit, and the median was used to provide the primary sky
subtraction. Total exposure times are given in Table~1.  The frames were then
registered, filtered to remove cosmic rays and artifacts, and then added. At
this point a second linear fit to the slit profile in the vicinity of the
quasar was used to remove any small residual sky. The observations were made
at the parallactic angle and flux calibrated using observations of white dwarf
standards scattered through the night. These standards were also used to
remove telluric absorption features in the spectra, including the various
atmospheric bands. The final extractions of the spectra were made using a
weighting provided by the profile of the white dwarf standards. Wavelength
calibrations were obtained using third-order fits to CuAr and Xe lamp
observations at the begining and end of each night, and the absolute
wavelength scale was then obtained by registering these wavelength solutions
to the night sky lines. The wavelengths and redshifts are given in the vacuum
heliocentric frame.

The quality of the data and the extraction is shown in Figure~1 for the
highest redshift QSO in the sample, SDSS~1044$-$0125, which also has one of the
lowest S/N in the sample.  The total exposure time was 5.75 hours.  The
spectrum is shown smoothed to $0.25~{\rm\AA}$\ and normalized in the region of
$1330 - 1370~{\rm\AA}$\ in the rest frame to a power-law continuum ($f_\lambda$)
with slope $-1.25$ (solid line).  The dotted lines show continua with slopes
$-0.75$\ and $-2$.  
For this range of slopes there is a $\pm 18$~\% range in the normalization at
$1075~{\rm\AA}$. 
We adopt the emission redshift of 5.745 for this quasar, which
Goodrich et al.\ (2001) measured from NIRSPEC spectroscopy of the \ion{C}{4}
emission line and associated absorption that they describe as BAL-like.
This associated absorption is only very weakly seen in the blue wing of the
\ion{Si}{4} line and the \ion{Si}{4} + \ion{O}{4}] emission gives a consistent
solution for the emission redshift.  The short vertical bars above the
spectrum show the resulting positions of the Ly$\beta$+\ion{O}{6}, Ly$\alpha$,
\ion{N}{5} and \ion{Si}{4} + \ion{O}{4}] emission lines.

%
%
\begin{inlinefigure}
\psfig{figure=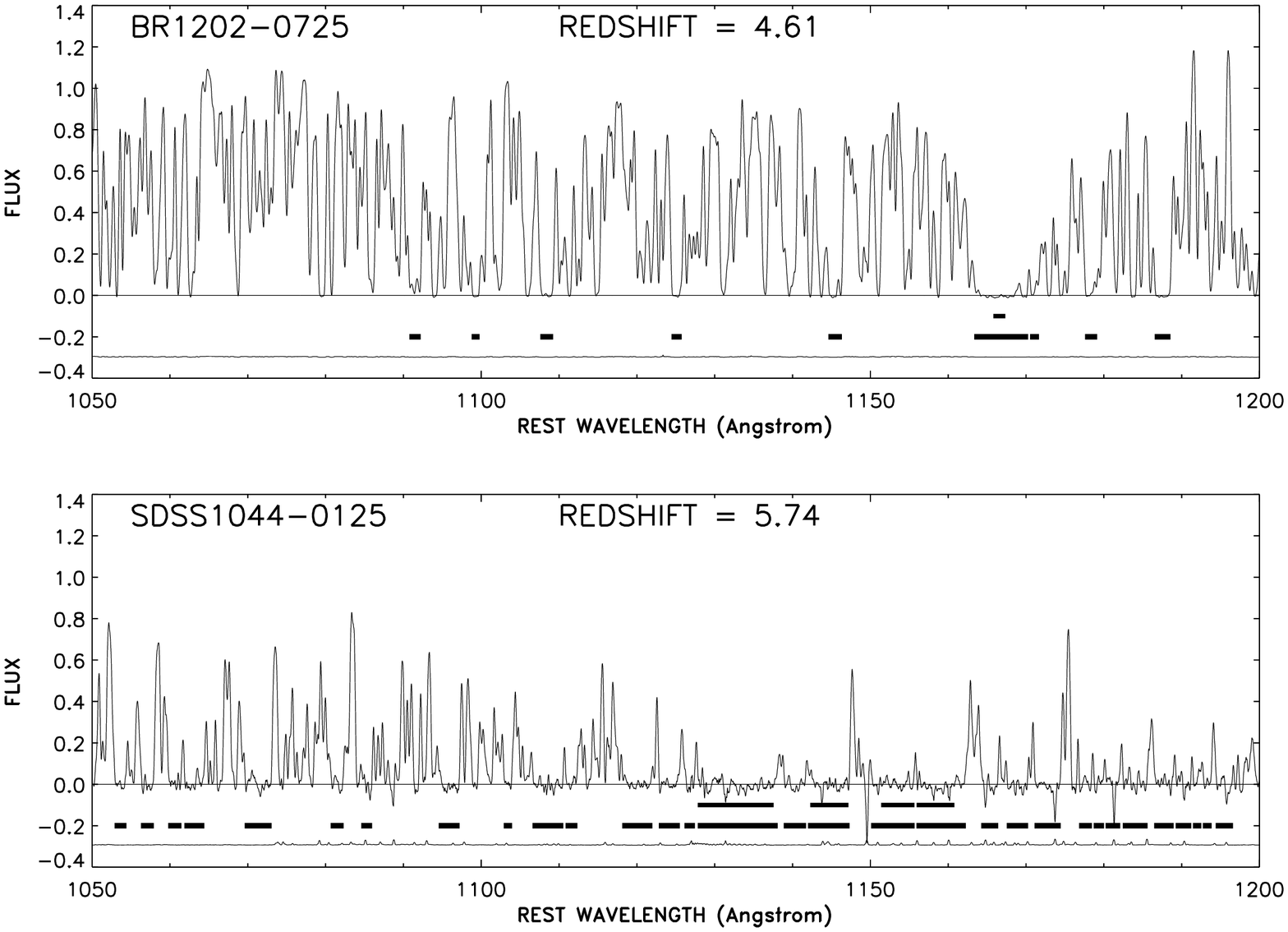,angle=90,width=7.0in}
\vspace{6pt}
\figurenum{2}
\caption{Portions of the spectra between rest wavelengths of $1050~{\rm\AA}$\
and $1200~{\rm\AA}$\ for (top panel) BRI~1202$-$0725 ($z_{em} = 4.61$) and
(bottom panel) SDSS~1044$-$0125 ($z_{em} = 5.74$), normalised to a power law
continuum with slope $-1.25$.  The formal statistical noise is shown in a panel
below each spectrum.  The black bars plotted imediately above the noise show
regions in the spectrum that have Ly$\alpha$\ optical depth, $\tau_{\rm
Ly\alpha} > 2.5$\ over a rest-frame width $> 1~{\rm\AA}$.  On the line above,
similar black bars mark regions of the spectra where the optical depth and
width criteria also hold at the position of Ly$\beta$\ in each spectrum.  }
\label{fig2}
\addtolength{\baselineskip}{10pt}
\end{inlinefigure}

The region of primary interest between $1050~{\rm\AA}$\ and $1200~{\rm\AA}$\
is shown in more detail in Figure~2 for SDSS~1044$-$0125 and for the lower
redshift and the more typical S/N spectrum of BRI~1202$-$0725, both of which
contain wide saturated regions which can be used to estimate the accuracy of
the sky subtraction.  The formal statistical noise derived for the spectra is
shown underneath the spectrum in each panel.  In general, however, systematic
residuals associated with strong night sky lines are a larger source of noise,
as can be clearly seen in SDSS~1044$-$0125.  In the base of the damped
Ly$\alpha$\ line at $z = 4.383$\ in BRI~1202$-$0725 we measure a mean
normalized flux of $-0.002$\ while in the dark region at $z = 5.286$\ in
SDSS~1044$-$0125 we measure $-0.009$, suggesting that the sky subtraction
leaves about 1\% residuals in the faintest quasars and smaller errors in the
brightest ones.  We take the larger value as our systematic error.

\section{Transmitted Fluxes and the Evolution of the Ionizing Parameter}

The simplest quantity that can be used to characterize the Lyman alpha forest
is the mean reduction which it introduces in the continuum.  Oke \& Korycansky
(1982) introduced the $D_A$\ index, defined as 
\begin{equation}
D_A = \left \langle 1 - {{F_{\nu}} \over {F_{\nu 0}}} \right \rangle
\end{equation}\smallskip\noindent
where $F_{\nu}$\ is the actual flux and $F_{\nu 0}$\ is the continuum, and the
average is over rest-frame wavelengths from $1050~{\rm\AA}$\ to
$1170~{\rm\AA}$, which makes this measurement over most of the Ly$\alpha$\
forest region.  We show this quantity in Figure~3 where, to make the most
direct comparison with the lower redshift points of Schneider, Schmidt \& Gunn
(1991) and Kennefick, Djorgovski \& de Carvalho (1995), we used a spectral
index of 0.9 in extrapolating the continuum, equal to the mean Schneider et
al.\ value.  (The choice of continuum can shift the points vertically by
offsets of up to 0.2).  We have also shown the values measured by Becker et
al.\ (2001) in the three of their four quasars that do not overlap with the
present sample.  The present data show a smooth connection between the high
and low redshifts, indicating the rapid drop in the transmitted flux as we
move to the higher redshifts.

%
%
\begin{inlinefigure}
\psfig{figure=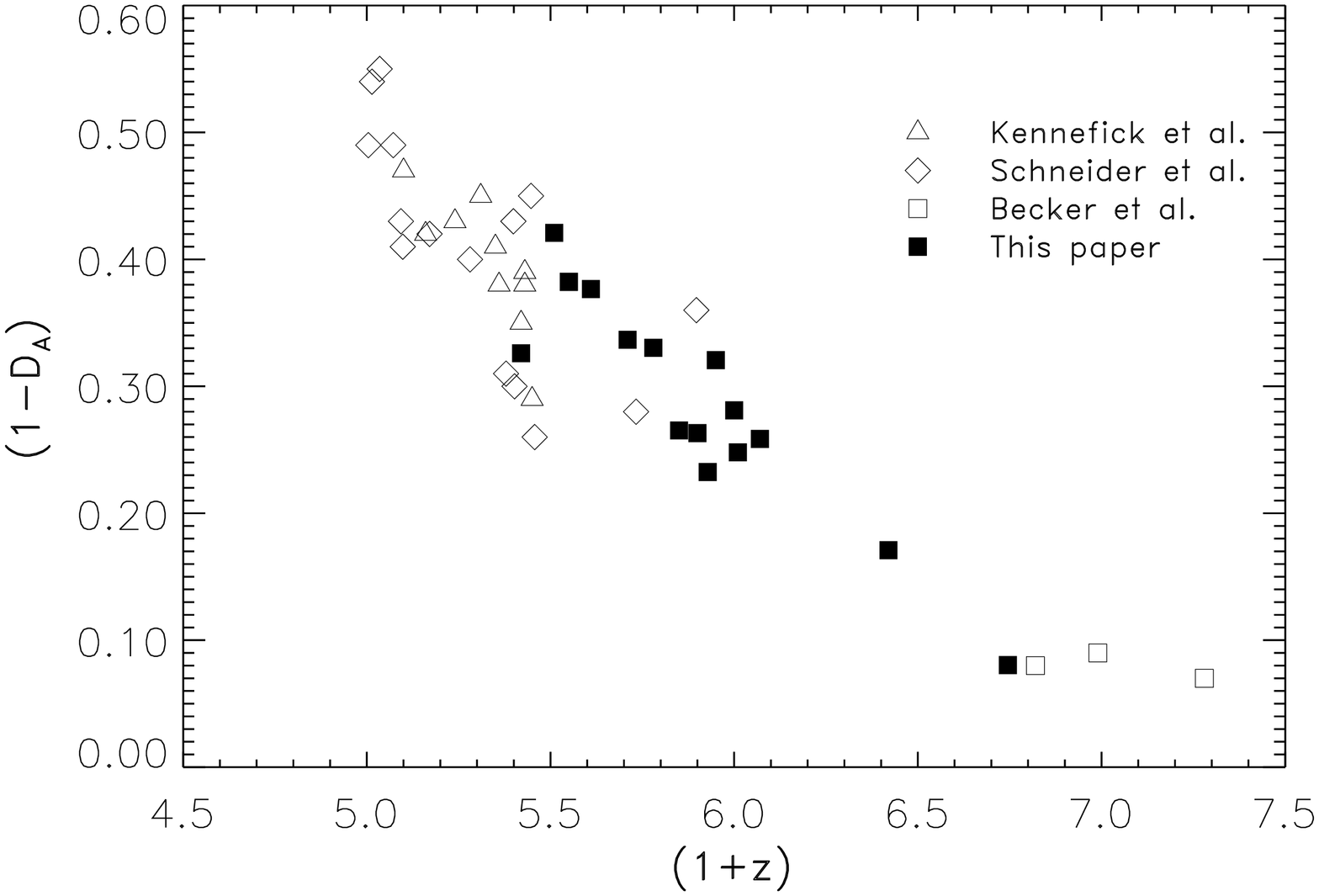,angle=90,width=4.in}
\vspace{6pt}
\figurenum{3}
\caption{Plot of $1 - D_A$, the Oke index, for the data of this paper
(filled squares), Becker et al.\ (2001; open squares), Kennefick, Djorgovski
\& de Carvalho (1995; open triangles) and Schneider, Schmidt \& Gunn (1991;
open diamonds).  }
\label{fig3}
\addtolength{\baselineskip}{10pt}
\end{inlinefigure}

The quality of the present spectra justifies a more detailed analysis, and for
each quasar transmitted fluxes blueward of the Ly$\alpha$\ emission were then
measured in four bins with rest wavelengths of 1150--1175~{\rm\AA},
1125--1150~{\rm\AA}, 1100--1125~{\rm\AA}\ and 1075--1100~{\rm\AA}.  For SDSS
1044$-$0125, these bins are illustrated as short horizontal bars below the
spectrum in Figure~1.  For the remainder of the paper we use the $-1.25$\
power law 
continuum from Figure~1.  The continguous wavelength bins used by Becker et
al.\ (2001) in calculating transmitted fluxes for this quasar are also shown
in Figure~1 with centers marked by the small squares.  It can be seen that the
Becker et al.\ bins extend considerably closer to the quasars' Ly$\alpha$\ and
Ly$\beta$\ emission, where the emission line flux might affect the
normalization, which is why we adopt the present intervals.  Measurements made
closer to the quasar redshift might also be more vulnerable to effects
associated with the quasar itself, as we discuss further below.  Some idea of
the possible errors in the transmitted fluxes in the shorter exposure Becker
et al.\ data can be gained from Figure~4, which is a comparison of the
transmitted fluxes in 1044$-$0125 given by Becker et al.\ with the fluxes
measured in the Becker et al.\ bins in the spectrum of Figure~1.  At these low
fluxes, the statistical errors in the present spectra are tiny, and even the
systematic errors from continuum fitting (see Figure~5 caption) are very
small.  Despite this, there are clearly substantial differences between the
Becker et al.\ points and the present data, which are most probably related
to the sky subtraction issue and to the lower S/N of the Becker et al.\ data.
The two highest wavelelength points are a factor of 1.7 lower than the Becker
et al.\ numbers.

The mean transmitted flux in these bins versus redshift is shown in Figure~5 
for the entire sample (filled squares).  The $1~\sigma$\ statistical error 
together with the continuum fitting uncertainty is
shown as a vertical error bar which is, however, sometimes smaller than the
plotting symbol, and the width of the bin is shown by the horizontal line.
The continuum fitting errors strongly dominate, particularly at higher
transmission.   We
also show (open squares) the transmitted fluxes measured by Becker et al.\ in
their three quasars that do not overlap with the present sample.  The overall
sample shows a smooth decline with redshift, albeit with a substantial
scatter.  A least squares fit to a power law of the form $F \propto (1+z)^x$\
gives $x = -7.1$\ if we use only the present data (solid line in Figure~5)
and $-8.1$\ if we include the Becker et al.\ data (dashed line).

%
\begin{inlinefigure}
\psfig{figure=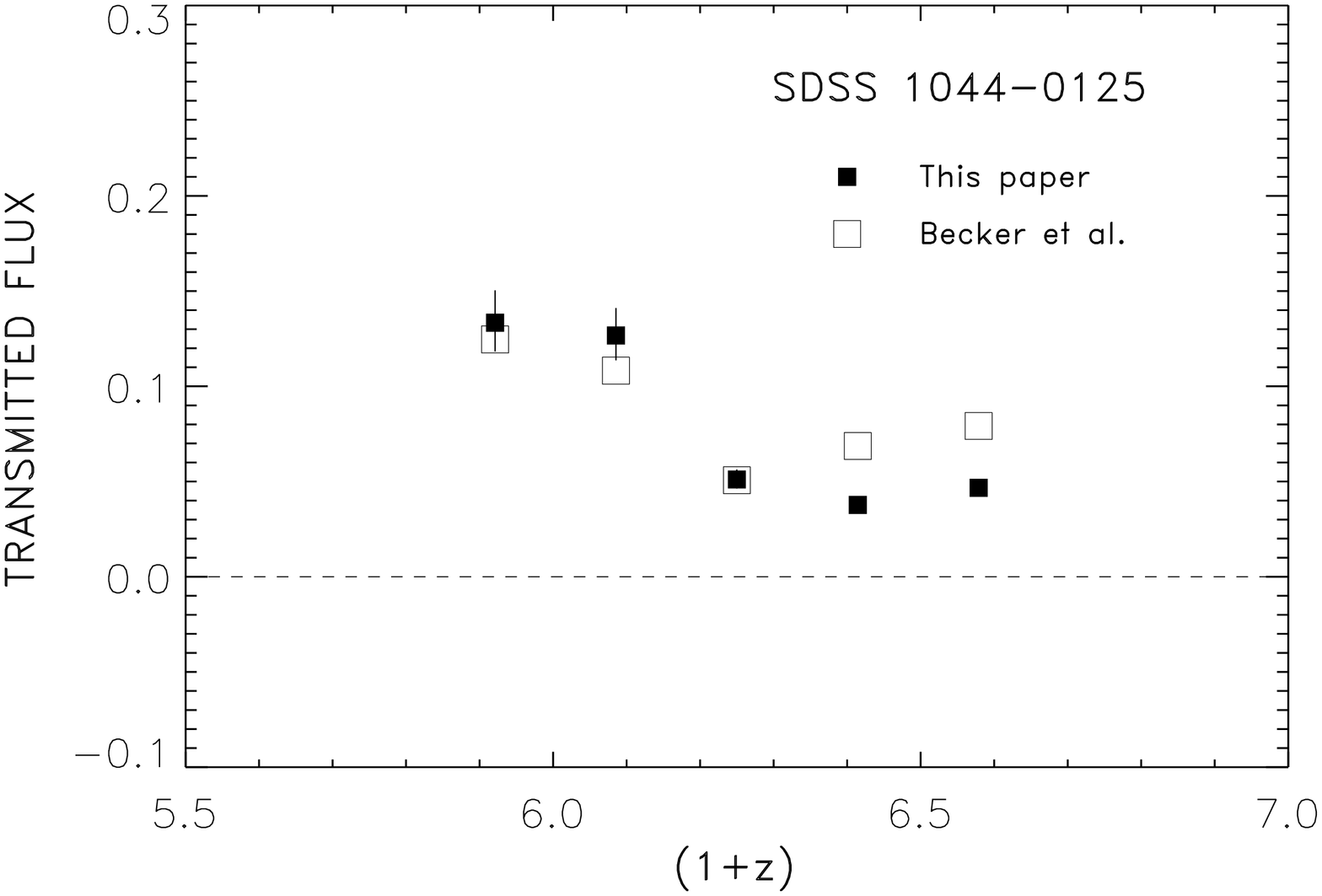,angle=90,width=3.7in}
\vspace{6pt}
\figurenum{4}
\caption{Comparison of the transmitted flux blueward of ${\rm Ly}\alpha$\ in
the quasar SDSS1044$-$0125 determined by this paper (filled squares) and by
Becker et al.\ (2001; open squares).   The error bars are described in
Figure~5.
}
\label{fig4}
\addtolength{\baselineskip}{10pt}
\end{inlinefigure}

%
%
\begin{inlinefigure}
\psfig{figure=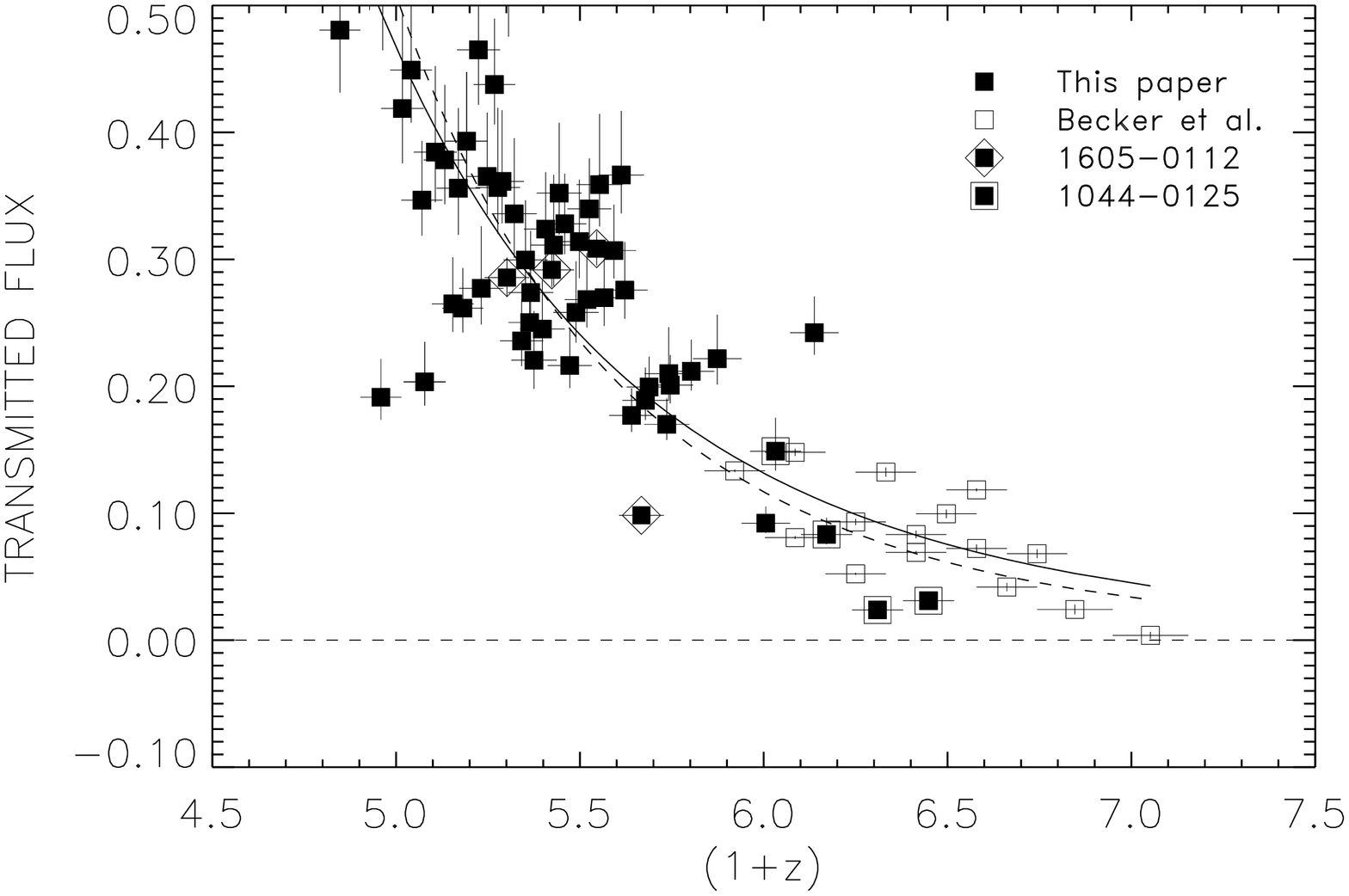,angle=90,width=6.0in}
\vspace{6pt}
\figurenum{5}
\caption{ Plot of transmitted flux vs (1+z) for the data of this sample
(filled squares) and of Becker et al.\ (1001; open squares).  (Statistical)
error bars from Becker et al.\ are shown as vertical lines.  For the present
sample, the error bars are a combination of statistical errors and errors from
varying the continuum fit (see Figure~1); in some cases they are smaller than
the data boxes.  These errors are strongly dominated by the continuum fitting
error.  The transmitted fluxes of the possible BAL QSO SDSS1044$-$0125 and of
the BAL QSO SDSS1605$-$0122 are marked.  (The latter object is excluded from
the sample.)  The solid and dashed lines are fits of the form $F \propto
(1+z)^x$.  The solid line is fitted only to the present data, and has an index
of $-7.1$\ whereas the dashed line fit includes the Becker et al.\ data and
has an index of $-8.1$.  
}
\label{fig5}
\addtolength{\baselineskip}{10pt}
\end{inlinefigure}

At a given redshift, the scatter in the transmitted flux represents the
variance produced by different sampling of the density distribution function
(and possibly the variation in the ionization parameter $g$) in the wavelength
average over the particular bin.  However, some part of this scatter appears
to be associated with intrinsic quasar absorption.  Two of the quasars,
SDSS~1605$-$0112 and SDSS~1044$-$0125, show BALQSO behaviour.
SDSS~1605$-$0112 is a clear BALQSO and SDSS~1044$-$0125 has strong associated
lines that were described by Goodrich et al.\ (2001) and can also be seen in
Figure~1 in \ion{Si}{4}.  We have distinguished the transmitted flux from
these quasars in Figure~5, and in both cases, the longest wavelength points
appear anomalously low compared with the typical value.  It appears probable
that this is being caused by Ly$\alpha$\ opacity associated with the quasar.
This type of behaviour is clearly not uncommon, with 2 of the current sample
of 15 quasars falling into that category.  We do not yet know whether the
highest redshift quasars in the Becker et al.\ sample are of like type, which
could result in low transmission in the highest redshift point, and this is
clearly an issue that must be checked (but see Pentericci et al.\ 2001).  We
have (slightly arbitrarily) excluded SDSS 1605$-$0112 but not
SDSS~1044$-$0125, from the sample at this point, but the results are not
significantly affected by the inclusion or exclusion of either object.

We next formed the mean
transmitted flux averaged over the measured values from the full sample of
quasars in the given redshift bin (including the data of Becker et al.\ (2001)
for the three non-overlapping quasars) and also the
variance in the points, which we used to determine the uncertainty in the
mean.  The mean values and errors are summarized in Table~2 and shown as the
large filled diamonds in Figure~6, where we have plotted the mean optical
depth, $\bar\tau$, defined as $-\ln ({\rm mean\ transmitted\ flux})$, versus
$\ln (1+z)$\ so that the power law fits, which are again illustrated, appear
as straight lines.  We have also shown the individual data points.  The
evolution of the mean optical depth from $z = 4$\ to $z = 6$\ is well
represented by 
\begin{equation}
\bar\tau = 3.5 + 8.1 \ln \left ( {{1+z} \over {7}} \right )
\end{equation}\smallskip\noindent
which corresponds to a mean transmission relation of
\begin{equation}
F = 0.031 \, \left ({{1+z} \over {7}} \right )^{-8.1}
\end{equation}

%
%
\begin{inlinefigure}
\psfig{figure=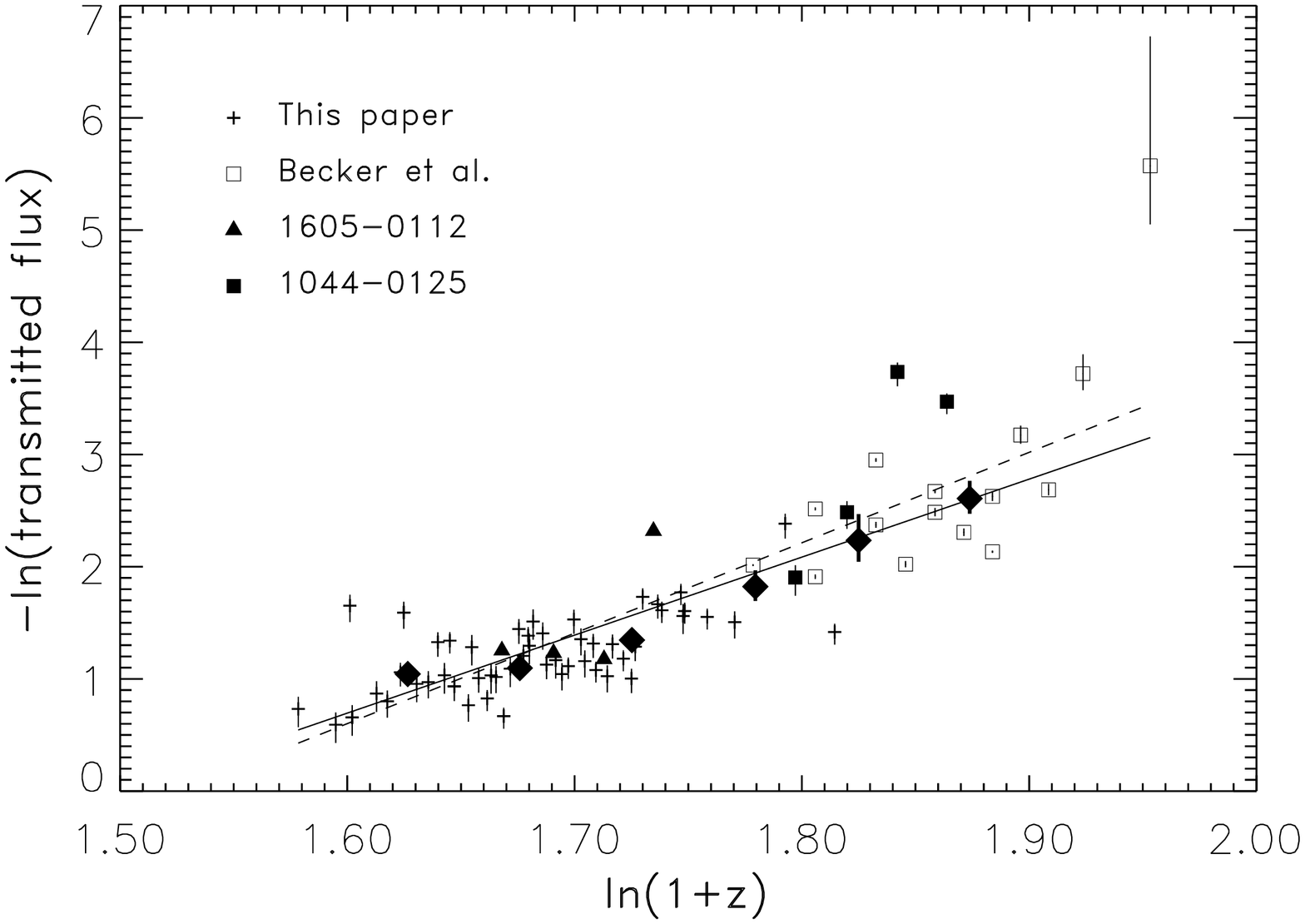,angle=90,width=4.5in}
\vspace{6pt}
\figurenum{6}
\caption{Plot of $-$ln(transmitted flux) vs ln(1+z) for the data of this
sample (crosses) and of Becker et al.\ (2001; open squares).  Error bars as
described in Figure~5 are shown only for the Becker et al.\ points.  Filled
squares show the values for the BAL QSO SDSS 1044$-$0125, and filled triangles
for the BAL QSO SDSS 1605$-$0122 (which is not included in the sample).  The
solid line is a power law of slope $7.1$\ fitted to the data of the current
sample, excluding the Becker et al.\ (2001) data, but including SDSS
1044$-$0125.  The dashed line is a power law of slope $8.1$\ fitted to all the
data except SDSS 1605$-$0122.  Filled diamonds show the average of all the
data except SDSS 1605$-$0122 in bins of width 0.05 in the log for $1.6 < {\rm
ln}(1+z) < 1.9$, with error bars showing the $\pm 1~\sigma$\ error in
the mean.  }
\label{fig6}
\addtolength{\baselineskip}{10pt}
\end{inlinefigure}

%
%
\begin{inlinefigure}
\psfig{figure=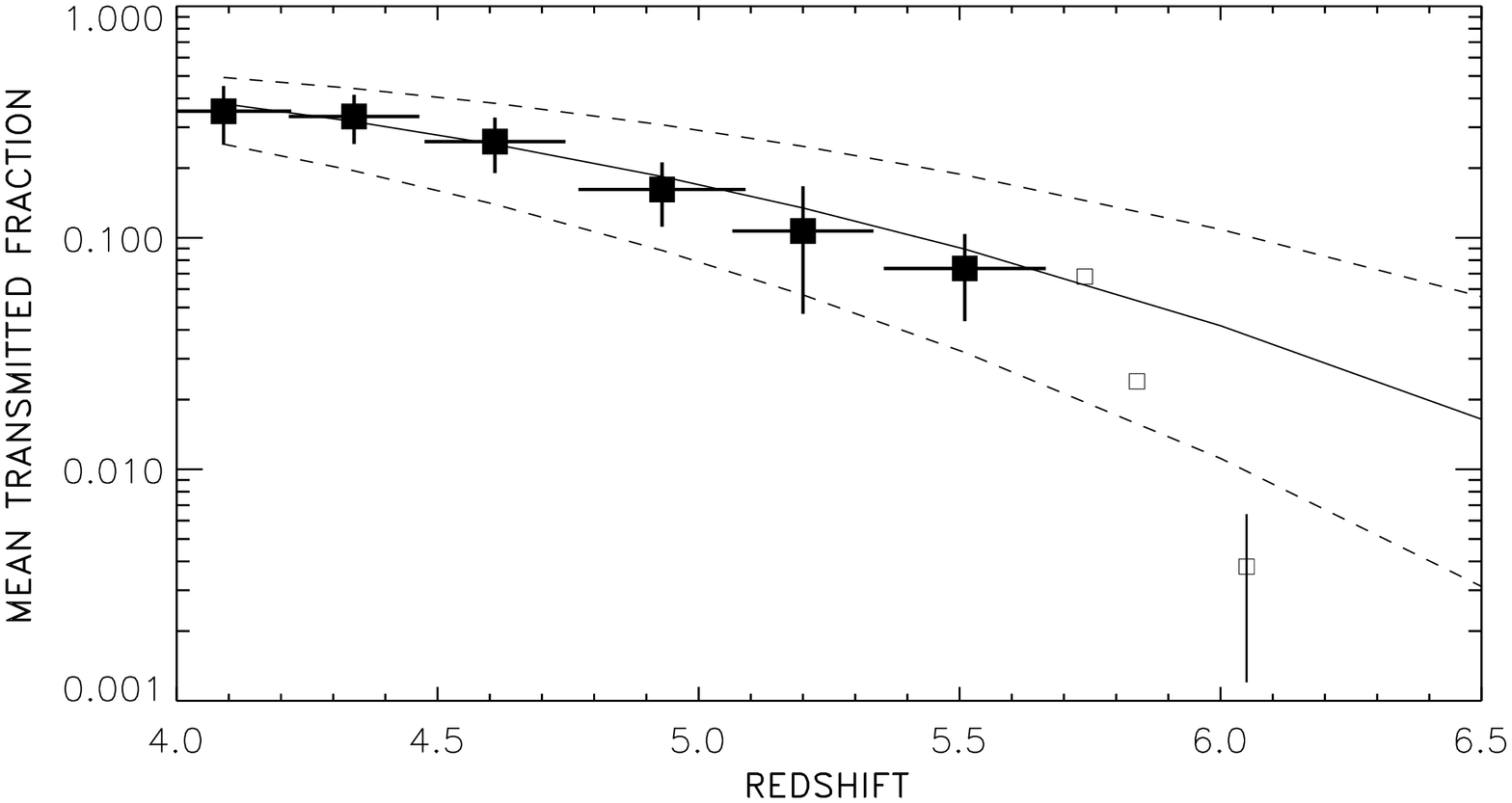,angle=90,width=4.5in}
\vspace{6pt}
\figurenum{7}
\caption{The mean transmitted flux as a function of redshift (filled squares)
compared with the expected transmission as a function of the ionization
parameter $g$\ (equation~17).  The model is shown for $g = 0.86$\ (solid line)
and $g = 0.43$\ and $g = 1.72$\ (dashed lines).  The error bars show the
variance in the individual measurements (rather than the error in the mean) to
illustrate the uncertainty (roughly a factor of 2) that would be associated
with measuring only a small number of transmissions at a given redshift.
Above $z = 5.7$\ we show the individual data points of Becker et al.\ (2001)
as small open squares.  For these Becker et al.\ points, the error bar is the
statistical uncertainty associated with the individual point.  }
\label{fig7}
\addtolength{\baselineskip}{10pt}
\end{inlinefigure}

We can now use equation(17) to translate the mean transmissions to ionization
rates.  The mean transmission and the variance of the transmission are
compared in Figure~7 with the expected evolution with $z$\ for an
ionization parameter $g = 0.86$\ which produces rough agreement with the data
points over the redshift range (solid line).  We also show the transmissions
expected when $g$\ is changed by a multiplicative factor of 2 in either
direction, which roughly corresponds to the variance, and provides a measure
of how much uncertainty would be expected from measuring $g$\ from a small
number of quasars. This can be used to assess the uncertainties in the
measurements from the individual Becker et al.\ (2001) data points at the
highest redshifts.

In Figure~8 we show the direct inversion to obtain $g$\ in each redshift bin
(filled points).  Here the error bars correspond to the error in the mean
transmission. We have also shown $g$\ determined by McDonald et al.\ (2000) at
lower redshift (open squares) and the values that would be determined from the
three individual Ly$\alpha$\ points in the Becker et al\ (2001) data at longer
redshift (small open diamonds).  As was shown in Figure~7, the present data
points are consistent with being constant as a function of redshift but the
combination with the lower redshift data points shows that the 
ionization parameter can be quite well represented by a power law of the form
\begin{equation}
g = 41 \, (1+z)^{-2.2}
\end{equation}\smallskip\noindent
which is shown as the solid line in Figure~8.  Only the $z = 4$\ data point
deviates substantially from this form and this is the most uncertain
point because of the continuum extrapolation.  In contrast to the results of
Cen \& McDonald (2001), whose findings were based on a small data sample, we
find no significant structure or peaks in the ionization function that might be
associated with the period just following reionization.  Instead, with the
single exception of the highest redshift Becker et al.\ (2001) point, all the
data are consistent with a smooth evolution to $z = 6$.  The possibility that
the reionization epoch is at $z \sim 6.1$\ stands or falls, therefore, on this
single data point at $z = 6.05$. 

$g$\ is related to the comoving production rate of ionizing photons,
$\dot{n}_I$, released to the IGM by the relation 
\begin{equation}
\dot{n}_I \, (1+z)^3 = {{\Gamma} \over {\sigma \, \Delta\lambda}}
\end{equation}\smallskip\noindent
(e.g.\ McDonald \& Miralda-Escud\'e 2001) where $\Delta\lambda$\ is the proper
distance between absorbers and $\sigma$\ is the mean cross section for
ionization.  Madau, Haardt \& Rees (1999) give $\Delta\lambda = 33~{\rm
Mpc}(1+z)^{-4.5}$\ and using this form, setting $\sigma = 2 \times
10^{-18}~{\rm cm}^{-2}$, and adopting the cosmology of section~2 so that
$\Gamma = 10^{-12}g~{\rm s}^{-1}$, we have
\begin{equation}
\dot{n}_I \approx 2 \times 10^{-19} (1+z)^{-0.7} \ \ {\rm cm^{-3}\ s^{-1}}
\end{equation}\smallskip\noindent
so that the comoving production rate of ionizing photons is varying only slowly
with redshift.  We can integrate equation~(23) to find the total comoving
number density of ionizing photons produced at redshift $z$,
\begin{equation}
n_I(tot) = 0.5\,H_0^{-1} \Omega_m^{-0.5} (1+z)^{-1.5} \dot{n}_I 
          \approx  5 \times 10^{-2} (1+z)^{-2.2} \ \ {\rm cm^{-3}} 
\end{equation}
\smallskip\noindent
which is more than sufficient to ionize the hydrogen density ($n_{H_0} = 2.6
\times 10^{-7}~{\rm cm^{-3}}$\ with the presently adopted parameters) out to
the highest observable redshifts, even allowing for the multiplicative factor
of several uncertainty in the ratio of $n_I(tot)$\ to $n_{H_0}$\ which is
present in (21) from the adopted cosmology and the equation of state.

%
%
\begin{inlinefigure}
\psfig{figure=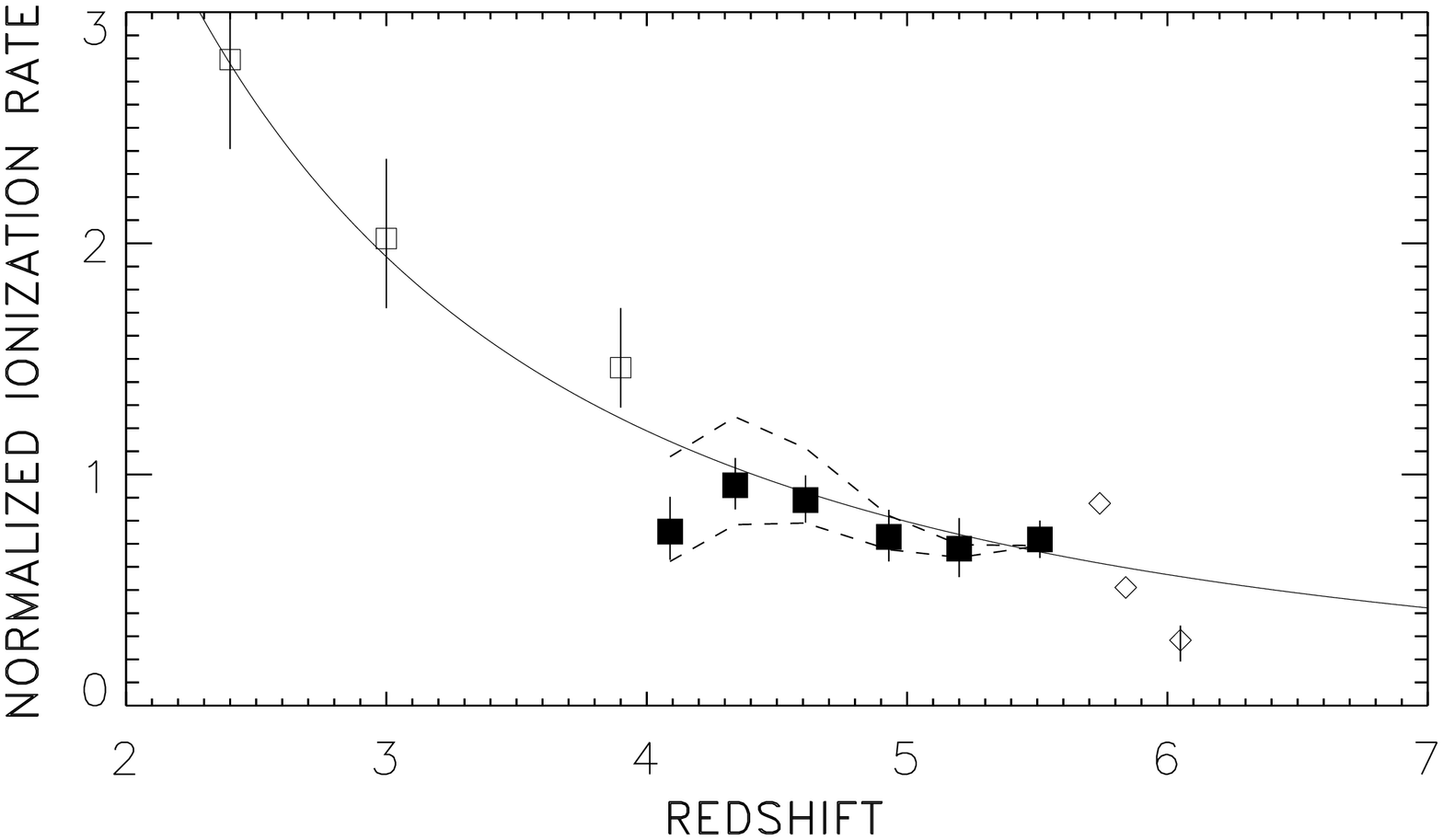,angle=90,width=4.5in}
\vspace{6pt}
\figurenum{8}
\caption{The evolution of the ionization parameter $g$\ as a function of
redshift computed from the mean transmissions using equation~15 (filled
squares).  The error bars correspond to errors in the mean.  The open squares
show the measurements of McDonald et al.\ at lower redshifts.  The small open
diamonds show the values measured from the individual Becker et al.\ (2001)
data points at higher redshift.  The solid line shows a fit of the form $g =
41 (1+z)^{-2.2}$\ which provides a reasonable description of the evolution of
the ionization parameter from $z = 2$\ to $z = 6$. The dashed lines show
the systematic errors which could be produced by the choice of the continuum
fit.
}
\label{fig8}
\addtolength{\baselineskip}{10pt}
\end{inlinefigure}

\section{The Distribution of Transmissions and the Incidence of Dark Gaps}

At redshifts $z > 5$\ is it no longer possible to carry out the kind of Voigt
profile analysis of the spectrum that has been used extensively at lower
redshifts (e.g.\ Hu et al.\ 1995; Kirkman \& Tytler 1997; Kim et al.\ 1997).
However, subject to the uncertainties in the continuum extrapolation used to
normalize the spectrum 
(primarily important at low optical depths), we can measure alternative
quantities to characterise the spectra and compare with models, such as the
distribution of the transmissions (e.g.\ Rauch et al.\ 1997) or the frequency of
gaps (e.g.\ Croft 1998).

The simplest quantity that can be used to quantify the spectra is the
cumulative distribution of transmissions, which primarily depends only on the
distribution of the neutral hydrogen density at a particular redshift and
contains little information about the velocity structure.  We have computed
this quantity from the present observations for three redshift intervals
[4.25,4.75], [4.95,5.45] and [5.45,5.95].  The first two redshift intervals
are chosen to match the LCDM models presented by McDonald \& Miralda-Escud\'e
(2001).  The observational results are shown as the crosses in Figure~9, where
low redshift is at the bottom of the figure and the highest redshift is at
the top.  We also show the model distributions at $z = 4.5$\ (squares) and $z
= 5.2$\ (diamonds) from the LCDM numerical calculations, with ionization
parameters chosen to match the observations.  (The ionization rates are in
reasonable agreement with the values computed in the previous section from the mean
transmissions.)  This model agrees very well with the data, as was also found
by Fan et al.\ (2001b).  We can
approximate the numerical models by combining equations~(11) and (12) and this
result is shown as the solid lines in Figure~9.  We have used this to extend
the comparison of the model to $z = 5.7$\ where it remains in good agreement
with the observations.

Cen \& McDonald used this LCDM model to assess the probability of seeing a
dark transmission region similar to that in the highest redshift Becker et
al.\ (2001) point.  Using the more stringent constraint from Ly$\beta$, they
found that there was a $2~\sigma$\ probability of seeing such a feature if
their $\Gamma_{-12}$\ obeyed the relation $\Gamma_{-12} \le 0.062$, which
corresponds to $g = 0.3$\ in the present notation.  This can be compared to
the value of $g = 0.6$\ expected at $z = 6.05$\ from equation~(19) and might
suggest the beginning of a substantial drop in $g$, consistent with the Becker
et al.\ (2001) interpretation.

However, one possible problem with the models could lie in the assumption of a
constant ionization rate and equation of state (subsumed here in the quantity
$g$) at a given redshift.  Spatially correlated variations in these quantities
could be present even at post-reionization epochs and could substantially
change the probability of seeing dark gaps.  Clearly, if the distribution of
the $g$\ parameter were too broad, it would significantly change the shape of
the transmitted fraction shown in Figure~9, but within the uncertainties in
the data and the continuum normalization there is considerable latitude for
such a possibility.  We illustrate this in Figure~10 in which we have again
used equations (11) and (12) to construct the cumulative probability of the
transmission being below the given value.  For redshifts 5.2 and 5.7 we show
the data and the cumulative probability with a single ionization parameter
$g$\ chosen to optimize the fit.  We then show two models, in one of which
half the points in the spectrum have an ionization rate $0.5g$ (dashed line)
and in the other of which half the points have an ionization rate $0.25g$\
(dotted line).  In each case we have adjusted the ionization range in the
remaining half of the points to optimize the fit.  At $z = 5.2$, the dotted
line is substantially too high at low transmissions and it is clear that a
spread in ionization rate of a factor of 4 would produce too many dark regions
in the spectra; however, factors of 2 variation are quite acceptable.  By $z
= 5.7$, as the number of dark regions increases, even a factor of 4 variation
is not ruled out.  If such variations are present then the probability of
seeing a dark gap such as that seen by Becker et al.\ rises substantially, and
their result can be understood without invoking reionization (though such
variations in the ionization and equation of state, if present, might be
related to post-reionization effects).

To proceed further on these lines would require extensive simulations and
consideration of the physical processes that might give rise to such
variations.  An alternative approach is to look at the distribution of dark
gaps and the distribution of dark regions in the present data which can, in
principle, provide a more sensitive diagnosis of whether there are spatially
underionized regions.  One simple parameterization
is the distribution of the widths of contiguous regions with
optical depth greater than or equal to a particular value (Croft 1998).  In
Figure~2 we show the contiguous regions satisfying the condition $\tau > 2.5$\
over rest-frame wavelength intervals $> 1~{\rm\AA}$\ in the quasar
BR~1202$-$0725 ($z_{em} = 4.61$) and in SDSS~1044$-$0125 ($z_{em} = 5.74$).
As would be expected from the increasing opacities, the frequency of such gaps
increases rapidly with redshift.  We have also shown the regions where the
same condition holds also at Ly$\beta$.  Only one region in BR~1202$-$0725
(the DLA at $z = 4.383$) appears, but even by $z = 5.3$, wide systems that are
dark in both Ly$\alpha$\ and Ly$\beta$\ are becoming common.  We illustrate
the distribution of the gaps in Figure~11 where we show the number of gaps
versus the wavelength per unit redshift interval, $dX \equiv \Omega_m^{-0.5}\,
(1+z)^{0.5} \, dz$.  Below $z = 5$, the distribution of gaps
varies fairly slowly but at $z > 5$\ the number of wide gaps increases
rapidly.  Again, further analysis requires more detailed simulations.

%
%
\begin{inlinefigure}
\psfig{figure=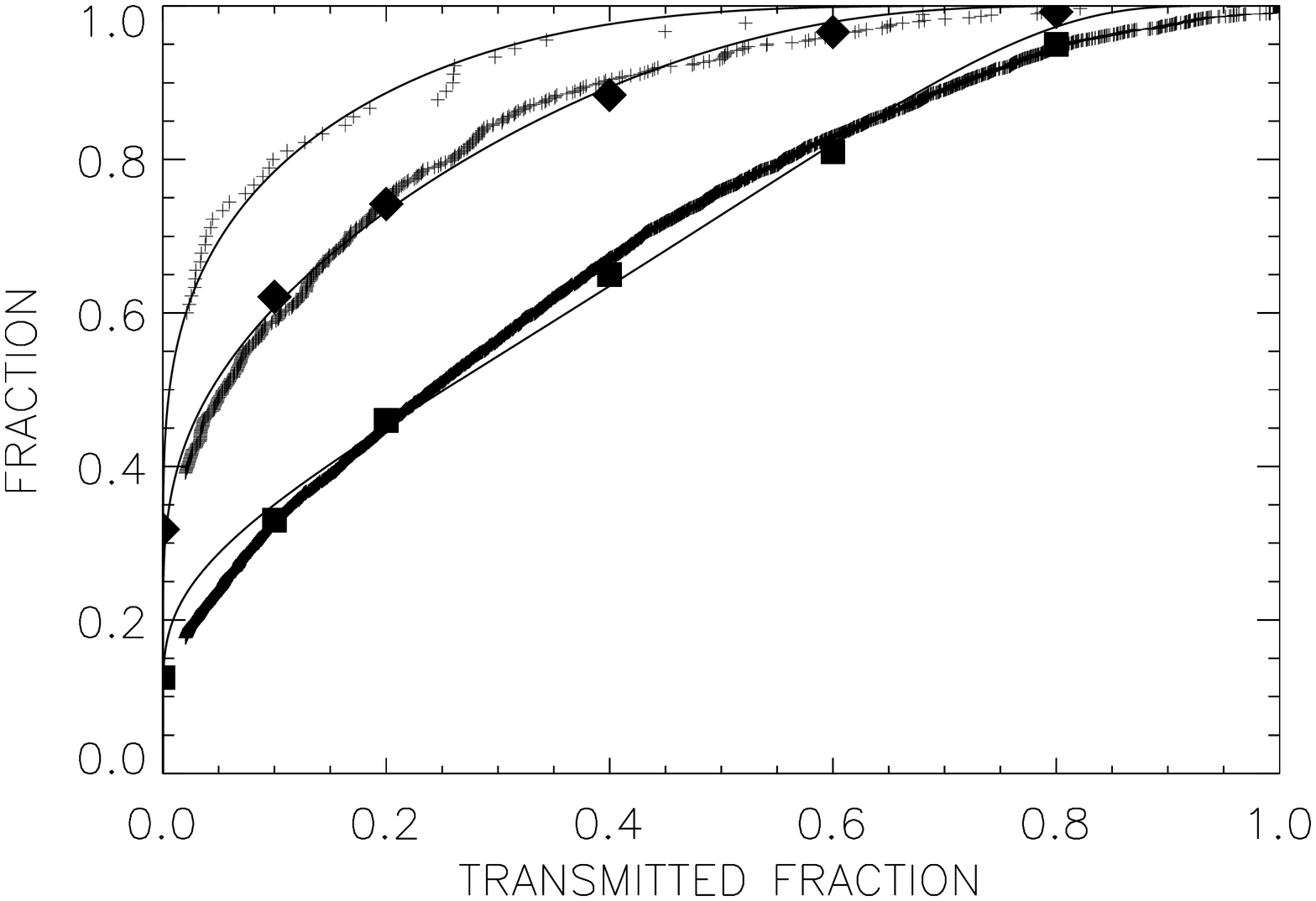,angle=90,width=6.0in}
\vspace{6pt}
\figurenum{9}
\caption{The crosses show the cumulative distribution of
the transmission fraction for three redshift
intervals ($z=4.25-4.75$, bottom) ($z=4.95-5.45$, middle)
and ($z=5.45-5.95$, top). The measured
values are compared with the results predicted
by the LDCM model of McDonald and Miralda-Escud\'e
at $z=4.7$\ (squares) and $z=5.2$\ (diamonds) where we
have used $\Gamma_{-12}=0.28$\ at the lower redshift
and 0.23 at the higher, in their units, corresponding
to $g=1.2$\ at $z=4.5$\ and $g=0.9$\ at $z = 5.2$. The model
is in excellent agreement with the data. We also show as solid lines the
values computed by combining the density distribution function of
equation~(12) with the opacity relation of equation~(11) for an equation of
state corresponding to $\beta = 1.75$.  The analytic approximation provides a
reasonable representation of the numerical models and the data at the two
lower redshifts, and matches well to the data at $z = 5.7$\ for a choice of $g
= 0.9$.  The exact shape at the high transmission end is sensitive to the
extrapolated continuum.  This error corresponds to a multiplicative rescaling
of the x-axis by a factor of up to 1.2.
}
\label{fig9}
\addtolength{\baselineskip}{10pt}
\end{inlinefigure}

%
%
\begin{inlinefigure}
\psfig{figure=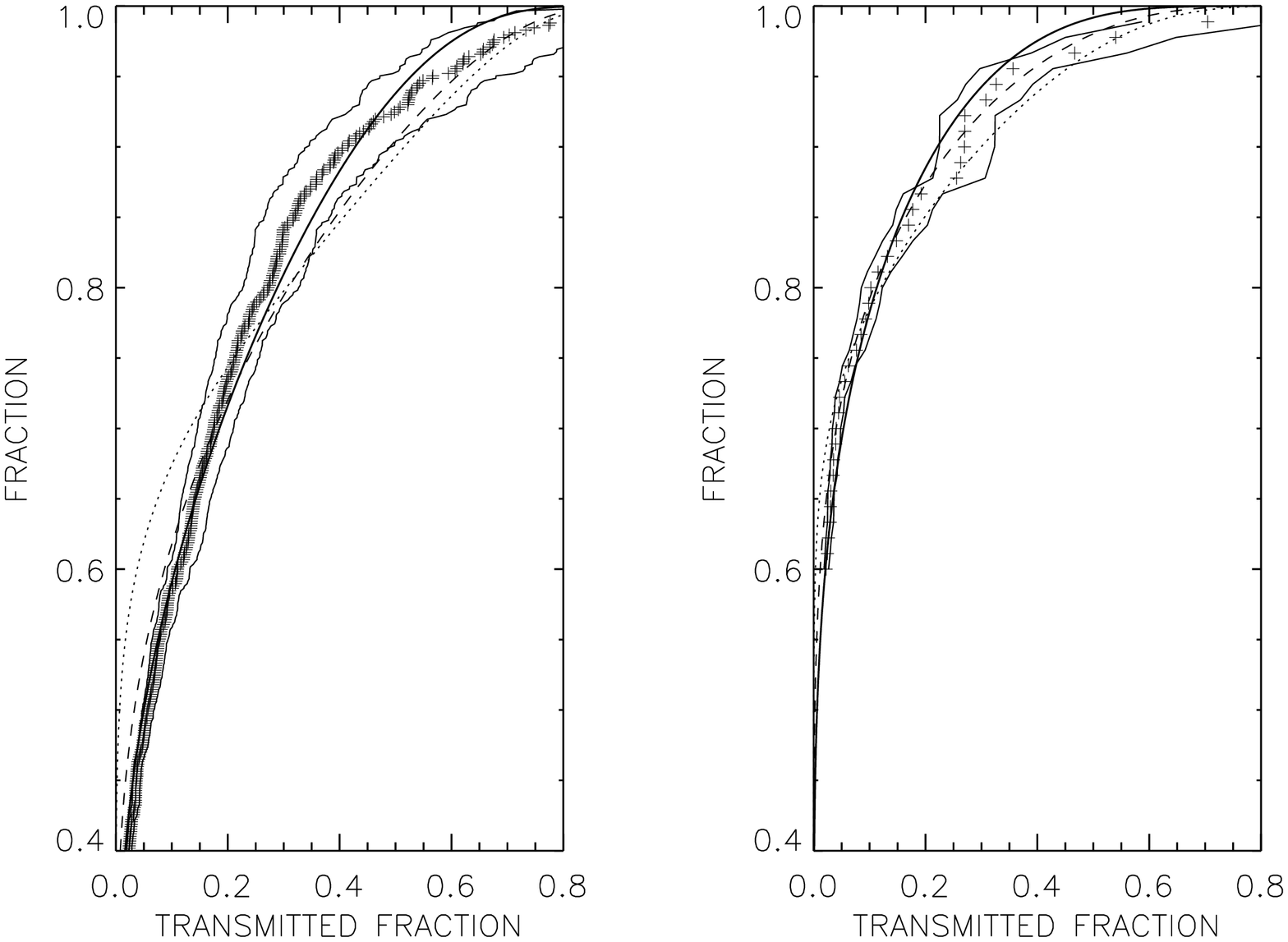,angle=90,width=5.0in}
\vspace{6pt}
\figurenum{10}
\caption{A comparison of the observed cumulative distribution function at $z =
5.2$\ (left panel) and at $z = 5.7$\ (right panel) with models in which $g$\
is constant and equal to the mean value given in Figure~9, and alternate
models in which there is variation in $g$\ about the mean.  At $z = 5.2$, the
dashed line shows a model where half the points have $g = 0.5$\ and the other
half have $g = 1.8$.  The dotted line shows the case where half have $g =
0.25$\ and half have $g = 1.9$.  For $z = 5.7$, the dashed line corresponds to
a similar equal split between $g = 0.4$\ and $g = 1.4$, and the dashed line an
equal split between $g = 0.2$\ and $g = 1.6$. 
}
\label{fig10}
\addtolength{\baselineskip}{10pt}
\end{inlinefigure}

%
%
\begin{inlinefigure}
\psfig{figure=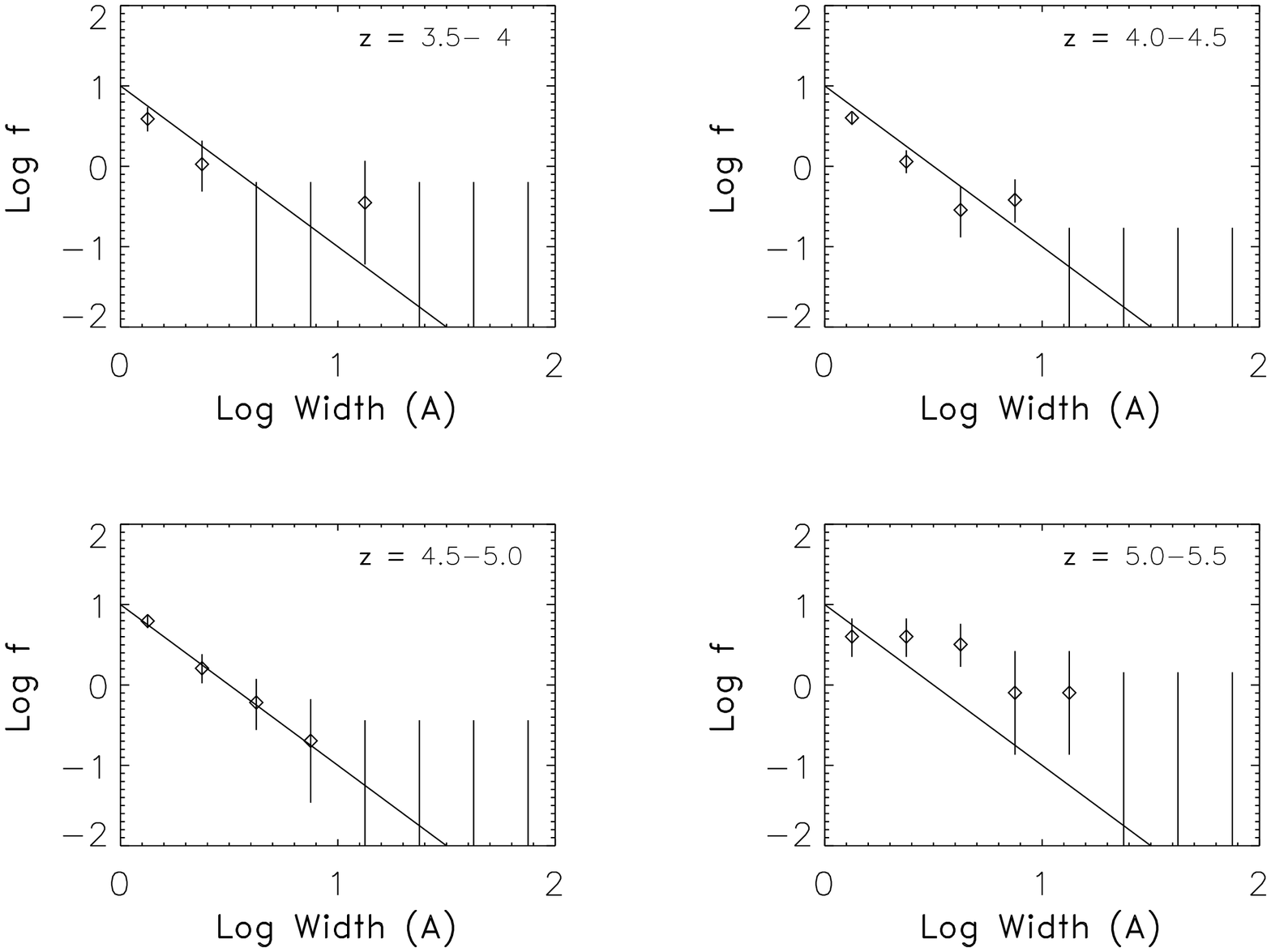,angle=90,width=6.0in}
\vspace{6pt}
\figurenum{11}
\caption{ Distribution of the rest-frame widths, in Angstroms, of 'gaps' with
optical depth $> 2.5$\ between
$1050~{\rm\AA}$\ and $1200~{\rm\AA}$\ (rest-frame) for the quasar sample of
pthis paper (excluding the Becker et al. sample).  $f$ is the number of gaps in
each bin per unit redshift path.  The bin size is $10^{0.25}$\ and the error
bars are $\pm 1 \sigma$\ based on the number of points in each bin.  The solid
line is a fiducial power law fitted to the $z = 4.5 - 5$\ distribution. }
\label{fig11}
\addtolength{\baselineskip}{10pt}
\end{inlinefigure}

\section{Metal Lines}

At $z < 4.5$\ the mean metal content of damped Lyman alpha systems (DLAs)
varies only slowly (Prochaska, Gawiser \& Wolfe 2001) and, while there is a
substantial dispersion in metallicity, all systems generally have metals with
a minimum metallicity relative to solar of $\sim 2 \times 10^{-3}$.  Since we
can measure \ion{C}{2} lines to column densities of $\sim 10^{13}~{\rm
cm}^{-2}$\ even in spectra with S/N similar to that of SDSS~1044$-$0125, if
these metallicites were maintained to higher redshift we should be able to
detect metal absorption to high column density neutral regions with $N({\rm
H~I}) \ge 10^{19}~{\rm cm}^{-2}$\ and study the kinematic structure and
velocity distribution of the gas.

Motivated by this idea, we have taken the list of 'dark gaps' with rest-frame
Ly$\alpha$\ widths $> 4~{\rm\AA}$\ from the sample developed in the previous
section (Table~3) and searched at each position for associated metal
lines. For each system we show in Figure~12 the Ly$\alpha$\ profile and the
corresponding lines of \ion{C}{2} ($1334~{\rm\AA}$), \ion{C}{4}
($1548~{\rm\AA}$), \ion{Si}{2} ($1526~{\rm\AA}$), and \ion{Si}{4}
($1393~{\rm\AA}$).  The longer wavelength member of the \ion{C}{4} doublet at
$1550~{\rm\AA}$\ can also be seen.  At $z < 5$, all of the gaps chosen in this
way correspond to conventional DLAs whose Lyman series profiles can be fitted
to the damping wings to obatin $N({\rm H~I})$.  The accuracy of the
measurements is about 0.2~dex and where there are previous measurements of the
same system (Prochaska et al.\ 2001 and references therein) the measured
values agree at this level.  We summarize these results in Table~3.  At higher
redshift ($z > 5$) the line blending makes this type of measurement extremely
difficult.  For the $z = 5.286$\ system in SDSS~1044$-$0125 we measure an
approximate $\log N({\rm H~I}) \sim 20.5$\ from Ly$\alpha$\ and Ly$\beta$.
The $z = 5.397$\ dark region has unsaturated regions at Ly$\beta$\ and clearly
has a lower column density.  For the remaining two $z > 5$\ systems, we give
upper limits on the column density based on the Ly$\alpha$\ and Ly$\beta$\
widths with $\tau > 2.5$\ using the equations of section~(2).

For each region we have also measured the column density of \ion{Fe}{2},
usually based on the $1608~{\rm\AA}$\ line, and used this to obtain the value
of [Fe/H].  No metal lines are seen in the three highest redshift systems.
These results are also summarized in Table~3.  

%
%
\begin{inlinefigure}
\psfig{figure=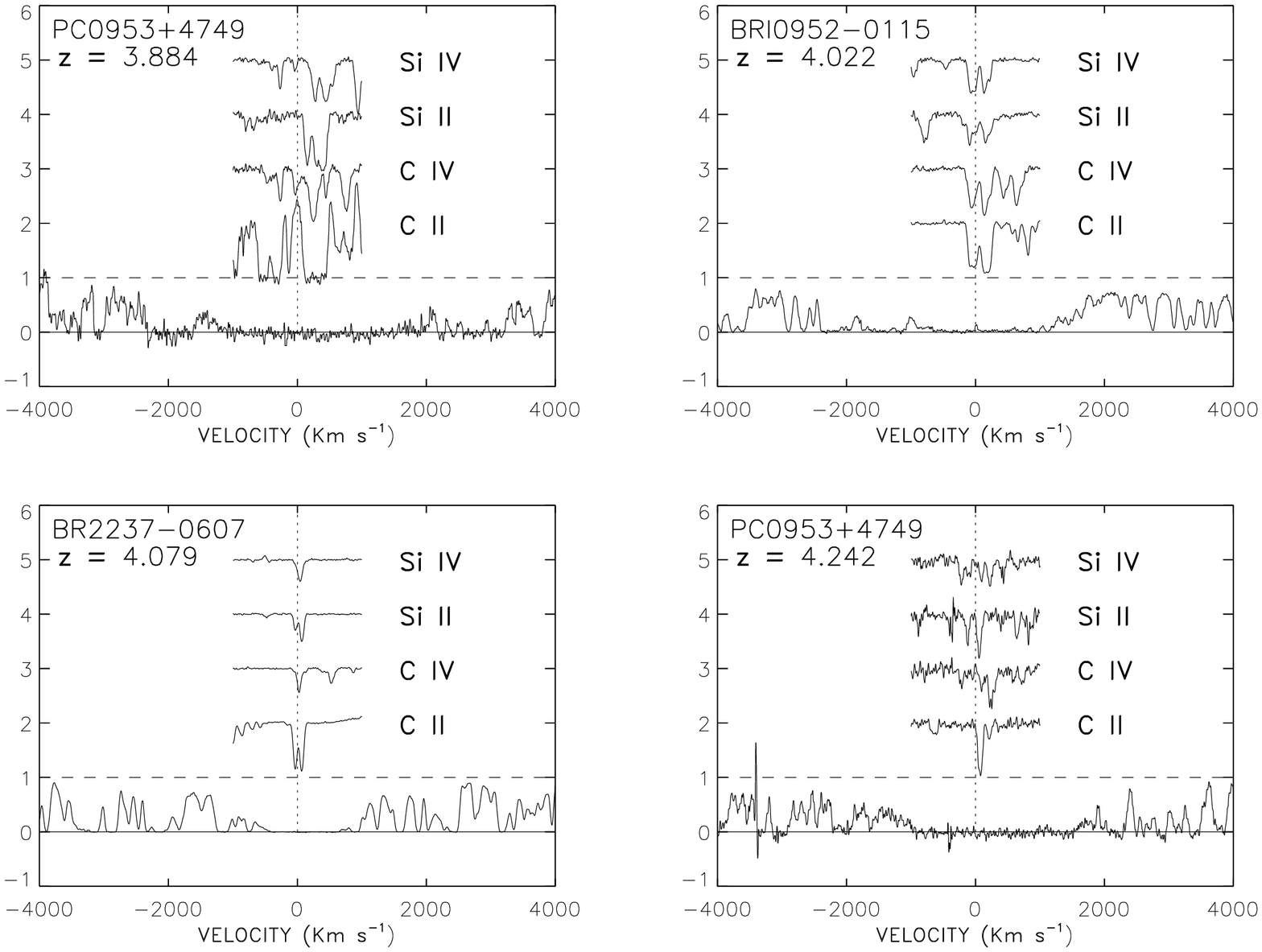,angle=90,width=6.0in}
\vspace{6pt}
\figurenum{12a}
\caption{Profiles of the absorption at Ly$\alpha$, C~II($1334~{\rm\AA}$),
C~IV($1548~{\rm\AA}$), Si~II($1526~{\rm\AA}$), and Si~IV($1393~{\rm\AA}$) for
the sample of Table~3.  The longer wavelength C~IV($1550~{\rm\AA}$) and 
C~II$^{\ast}$($1335~{\rm\AA}$) can also be seen in the C~IV and C~II profiles,
respectively.  The strong features seen at
the C~II position in SDSS~1737+5828 correspond to a lower redshift C~IV
doublet. 
}
\label{fig12a}
\addtolength{\baselineskip}{10pt}
\end{inlinefigure}

%
%
\begin{inlinefigure}
\psfig{figure=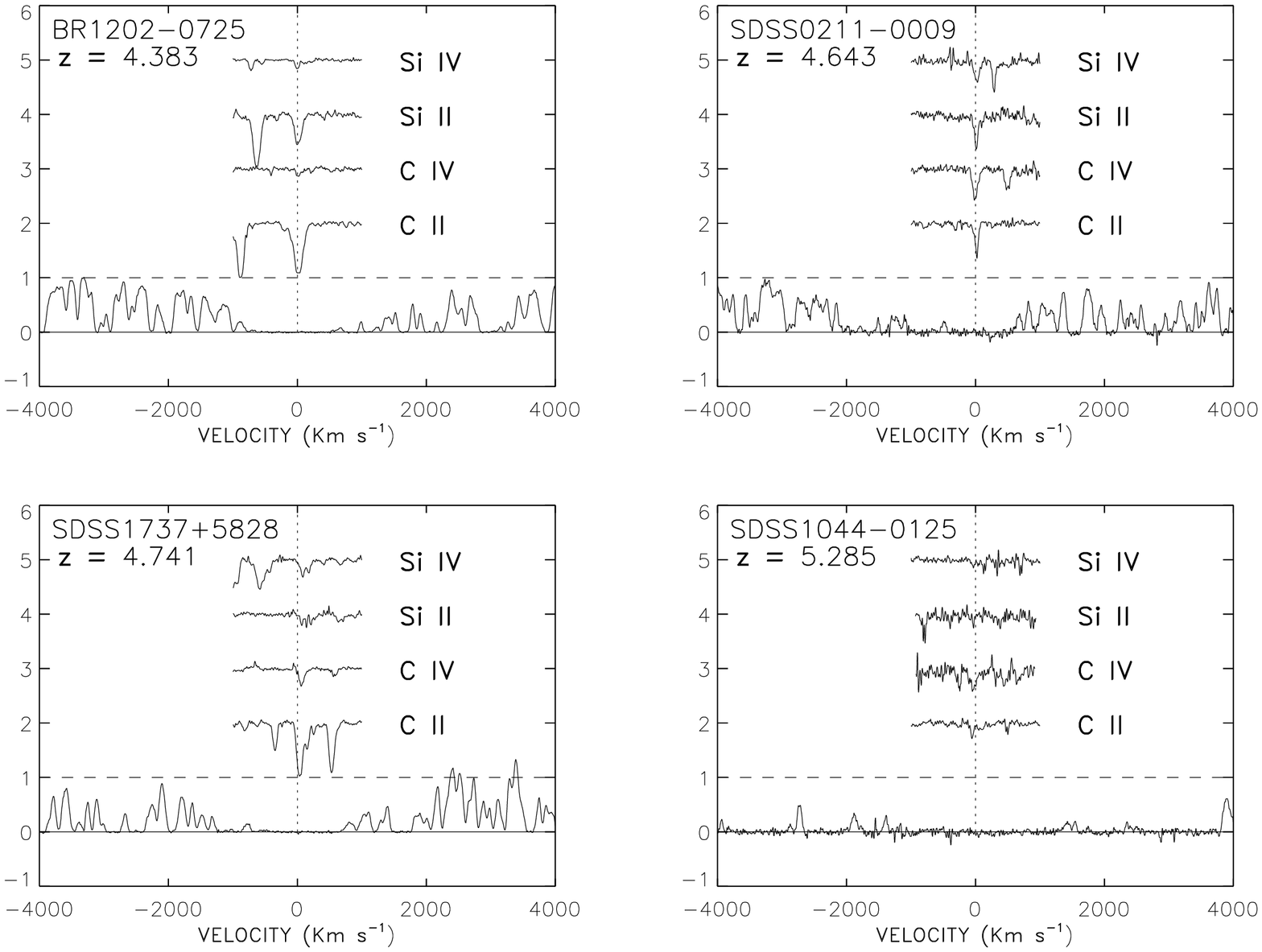,angle=90,width=6.0in}
\vspace{6pt}
\figurenum{12b}
\caption{Contd.
}
\label{fig12b}
\addtolength{\baselineskip}{10pt}
\end{inlinefigure}

%
%
\begin{inlinefigure}
\psfig{figure=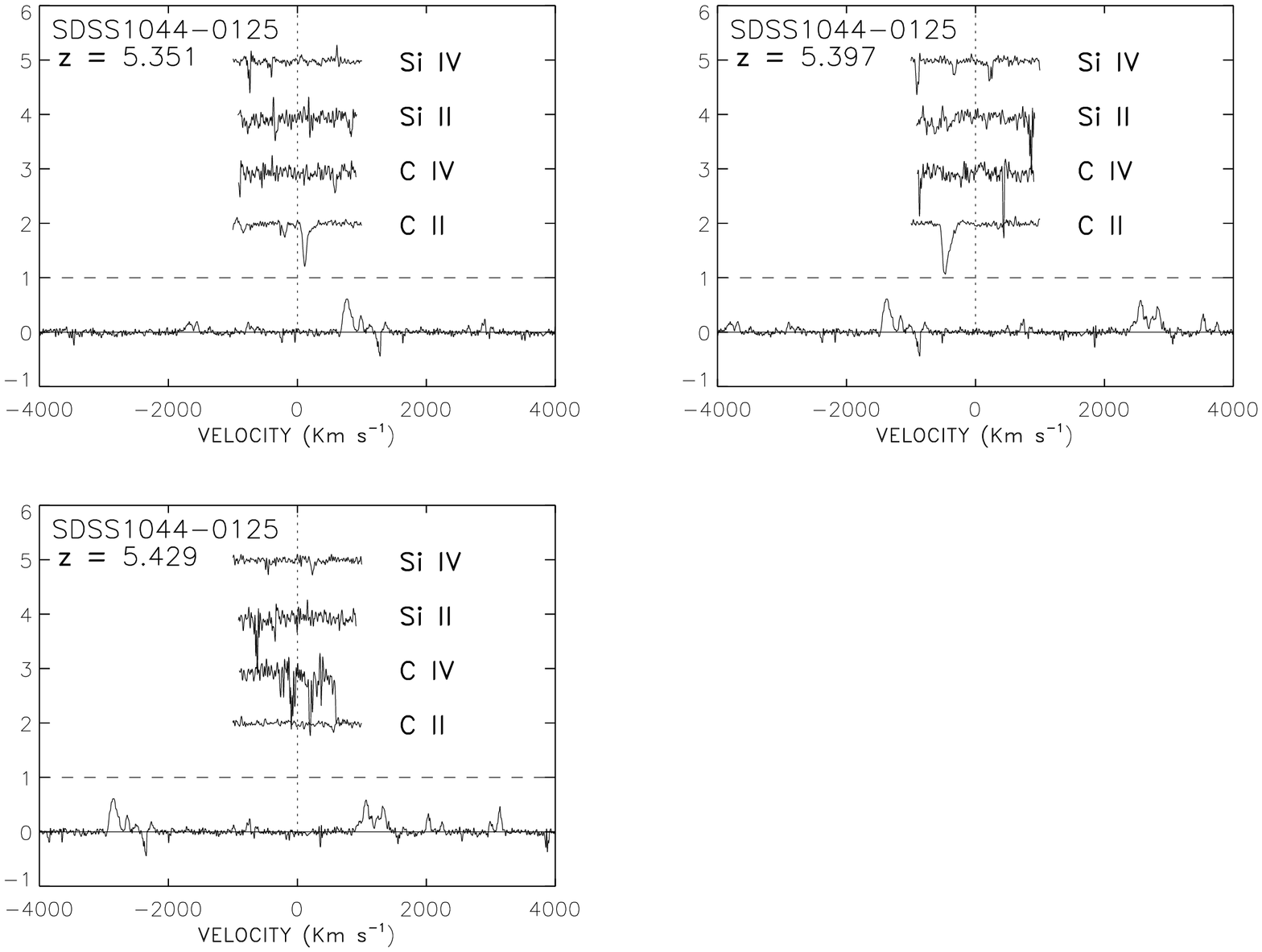,angle=90,width=6.0in}
\vspace{6pt}
\figurenum{12c}
\caption{Contd. 
}
\label{fig12c}
\addtolength{\baselineskip}{10pt}
\end{inlinefigure}

In Figure~13 we show [Fe/H] versus redshift (filled symbols) and we have also
shown the lower redshift measurements summarized in Prochaska et al.\ (2001;
open symbols).  Three of our measurements overlap with these previous results,
only one of which deviates by more than 0.2~dex: the present measurement of
the $z = 4.383$\ system in BR~1202$-$0725 is 0.3 dex higher than that of Lu et
al.\ (1996) which was, however, quite uncertain because of limited line
coverage.  The present measurement should be more accurate.  From Figure~13 we
can see that at these higher redshifts there is strong evidence that the
metallicity is beginning to fall.  The median metallicity at $z < 4$\ is
$-1.77$\ and only one of the 12 systems above $z = 4$\ has a metallicity
higher than this.  The median metallicity of the 9 systems with $4 < z < 4.5$\
is $-1.94$\ and for the 3 systems with $z > 4.5$\ it is $-2.61$.  If this
trend continues to higher redshift, many of the stronger neutral hydrogen
systems may not be seen in the metals and the systems that are seen may be
somewhat too weak for study of the kinematics.  However, even allowing for
these low metallicites, which result in weak lines not being seen, Figure~12
shows that these systems are kinematically very simple at the higher
redshifts.  All of the three metal systems at $z > 4.5$\ are strongly
dominated by a single narrow component.  In each case we have measured
$b$-values using the strongest unsaturated line of \ion{C}{2} or \ion{Si}{2}
available, obtaining $b = 19~{\rm km\ s}^{-1}$\ at $z = 4.64357$, $b = 19~{\rm
km\ s}^{-1}$\ at $z = 4.74314$\ in SDSS~1737+5828, and $b = 15~{\rm km\
s}^{-1}$\ at $z = 5.28538$\ in SDSS~1044$-$0125.  Taking into account the
broad instrumental profile, these $b$-values might best be treated as upper
limits, but irrespective of this point, it is clear that the high-$z$\ DLAs
are arising in kinematically quiescent regions, with line-of-sight turbulent
velocity dispersions of less than $10 - 14~{\rm km\ s}^{-1}$.

%
%
\begin{inlinefigure}
\psfig{figure=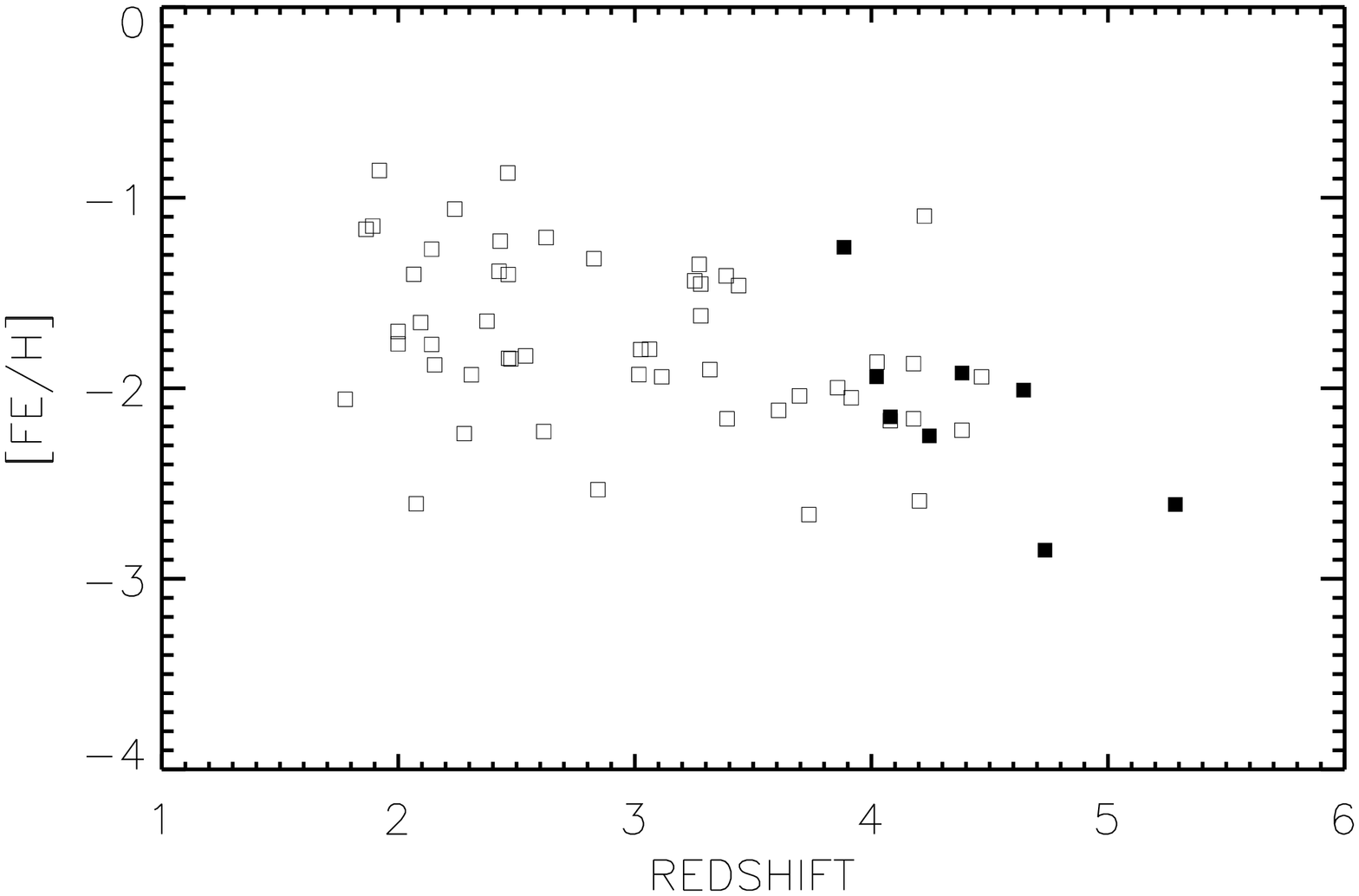,angle=90,width=6.0in}
\vspace{6pt}
\figurenum{13}
\caption{The metallicity relative to solar, based on the ratio of Fe~II to
H~I column densities.  The filled symbols show the present measurements and
the open symbols, the values summarized in Prochaska, Gawiser \& Wolfe (2001).
Three of the points (at $z = 4.024,\ 4.080\ and\ 4.383$) overlap between the
two samples.  The present value at $z = 4.383$\ is  higher than the value in
Prochaska et al.\ (see text). 
}
\label{fig13}
\addtolength{\baselineskip}{10pt}
\end{inlinefigure}

\section{Conclusions}

The primary result of the present paper is that between redshifts 4 and 6 the
Ly$\alpha$\ forest smoothly evolves from its behaviour at lower redshifts.
The presence of substantial 'dark gaps' in the Ly$\alpha$\ forest region is a
natural consequence of the increase in the gas density and the decrease in the
amplitude of density fluctuations and there are no signatures which require us
to lie in the immediate post-reionization universe at this epoch.  Over this
redshift range, the data are consistent with a constant ionization while, even
comparing with values measured at lower redshifts (e.g. McDonald et al.\
2000), the ionization rate at $z = 5$\ is still 40\% of that at $z = 3$.  This
in turn implies production  rates of ionizing photons which, extrapolated, could
easily ionize the intergalactic gas to much higher redshifts.

The one possibly compelling piece of evidence for the onset of reionization
just beyond $z = 6$\ is the wide very dark region seen at $z = 6.05$\ by
Becker et al.\ (2001) in the spectrum of SDSS~1030+0524.  The present data
are consistent with the LCDM model used by Cen \& McDonald (2001) to show
that, at the $2~\sigma$\ confidence level, the ionization rate at this point
must have dropped by a factor of 2 from that at a redshift just less than 6.
However, the data are also consistent with models in which there is a large
variation in the ionization parameter about the mean at any given redshift,
and in this case the probablility of seeing such a point becomes high enough
to be consistent with the extrapolation from lower redshift.  We have also
noted that 2 out of the 15 quasars show BAL QSO bahaviour and that this results
in a substantial increase in the Ly$\alpha$\ opacity near the quasar.

Finally, we have measured the metallicity of the gas in the damped Ly$\alpha$\
systems in the quasars and show that the metallicity in these systems is
beginning to drop at $z = 5$.  The kinematic structure of the metal lines
shows that, at these redshifts, the neutral hydrogen arises in single rather
narrow regions with line-of-sight velocity dispersions of $10-14~{\rm km\
s}^{-1}$. 

\acknowledgments 
We would like to thank Patrick McDonald, Renyue Cen, Xiaohui Fan and Bob
Becker for supplying tabular versions of data from their papers and/or their
numerical models, and Patrick McDonald, Michael Strauss and Julia Kennefick
for very useful comments on the text.  This research was supported by the
National Science Foundation under grants AST 96-17216 and AST 00-98480.

\begin{deluxetable}{lccc}
\tablewidth{300pt}
\tablecaption{Observations \label{tbl:1}}
\tablehead{
\colhead{Quasar} & \colhead{Mag.} & \colhead{Expo. (hrs)} 
& \colhead{$z_{em}$} } 
\startdata
SDSS 0211$-$0009  & 20.0  & 6.0  & 4.90  \nl
SDSS 0231$-$0728  & 19.2  & 3.0  & 5.42  \nl
SDSS 0338+0021    & 20.0  & 7.25 & 5.01  \nl
BRI 0952$-$0115   & 18.7  & 4.25 & 4.42  \nl
BR 1033$-$0327    & 18.5  & 2.0  & 4.51  \nl
SDSS 1044$-$0125  & 19.7  & 5.75 & 5.74  \nl
SDSS 1204$-$0021  & 19.1  & 3.0  & 5.07  \nl
BR 1202$-$0725    & 18.7  & 3.0  & 4.61  \nl
SDSS 1321+0038    & 20.1  & 2.0  & 4.71  \nl
SDSS 1605$-$0122  & 19.4  & 3.75 & 4.93  \nl
WFSJ 1612+5255    & 19.9  & 1.5  & 4.95  \nl
SDSS 1737+5828    & 19.3  & 7.5  & 4.85  \nl
SDSS 2200+0017    & 19.1  & 5.3  & 4.78  \nl
SDSS 2216+0013    & 20.3  & 3.0  & 5.00  \nl
BR 2237$-$0607    & 18.3  & 11.6 & 4.55  \nl
\enddata
\end{deluxetable}

\begin{deluxetable}{lcccc}
\tablewidth{300pt}
\tablecaption{Mean Transmissions \label{tbl:2}}
\tablehead{
\colhead{$\langle z \rangle$} & \colhead{\# points} 
& \colhead{$\langle F\rangle$} 
& \colhead{$\sigma$} & \colhead{$\sigma_{\rm mean}$} }
\startdata
 4.09 &  15 & 0.352 & 0.103  &  0.027  \nl
 4.34 &  20 & 0.334 &  0.088 &  0.020  \nl
 4.61 &  15 & 0.260 &  0.065 &  0.017  \nl
 4.93 &   5 & 0.162 &  0.049 &  0.022  \nl
 5.20 &   8 & 0.107 &  0.063 &  0.022  \nl
 5.51 &   7 & 0.074 &  0.028 &  0.011  \nl

\enddata
\end{deluxetable}

\begin{deluxetable}{lcccccc}
\tablewidth{450pt}
\tablecaption{DLA Ly$\alpha$\ Widths, Column Densities and  Metallicities\label{tbl:3}}
\tablehead{
\colhead{Quasar} & \colhead{$z_{abs}$} 
& \colhead{$dW(Ly\alpha)$\tablenotemark{(a)}} 
& \colhead{$z_{metal}$} 
& \colhead{$N(H~I)$}  & \colhead{$N(Fe~II)$}  & \colhead{[Fe/H]} } 
\startdata

SDSS 1044$-$0125 & 5.429 &  6.3 &  0.0   & $< 20.70$ &  0.00 & 0.00 \nl
SDSS 1044$-$0125 & 5.397 &  5.6 &  0.0   &  0.00     &  0.00 & 0.00 \nl
SDSS 1044$-$0125 & 5.351 &  5.3 &  0.0   & $< 20.55$ &  0.00 & 0.00 \nl
SDSS 1044$-$0125 & 5.286 & 10.3 &  5.284 & (20.50)&13.34& $(-2.65)$ \nl
SDSS 1737+5828   & 4.742 &  5.9 &  4.743 &  20.65  & 13.30 &$-2.85$ \nl
SDSS 0211$-$0009 & 4.643 &  4.1 &  4.644 &  20.00  & 13.49 &$-2.00$ \nl
BR 1202$-$0725   & 4.383 &  6.9 &  4.384 &  20.60  & 14.09 &$-2.00$ \nl
PC 0953+4749     & 4.243 &  8.0 &  4.244 &  20.70  & 13.95 &$-2.25$ \nl
BR 2237$-$0607   & 4.079 &  5.6 &  4.080 &  20.55  & 13.88 &$-2.15$ \nl
BRI 0952$-$0115  & 4.023 &  7.9 &  4.023 &  20.65  & 14.21 &$-1.95$ \nl
PC 0953+4749     & 3.884 &  9.8 &  3.883 &  20.85  & 15.09 &$-1.25$ \nl

\enddata

\tablenotetext{(a)}{Rest-frame Ly$\alpha$\ width for $\tau > 2.5$.}

\end{deluxetable}

\end{document}